%% file: main_arxiv.tex
\documentclass[10pt,A4paper,twocolumn,aps,pra, superscriptaddress,longbibliography, nofootinbib]{revtex4-2}

\input{preamble}

\begin{document}

\title{\puttitle}
\author{~}
\date{December 18, 2024}

\begin{abstract}

Quantum error correction~\cite{shor1995scheme, gottesman1997stabilizer, dennis2002topological, kitaev2003fault, terhal2015quantum, Litinski2019a} is essential for bridging the gap between the error rates of physical devices and the extremely low logical error rates required for quantum algorithms. 
Recent error-correction demonstrations on superconducting processors~\cite{Krinner2022, Zhao2022, Ye2023, Acharya2023, Sundaresan2023, Acharya2024, Ali2024} have focused primarily on the surface code~\cite{fowler2012surface}, which offers a high error threshold but poses limitations for logical operations.
In contrast, the color code~\cite{Bombin2006} enables much more efficient logic, although it requires more complex stabilizer measurements and decoding techniques.
Measuring these stabilizers in planar architectures such as superconducting qubits is challenging, and so far, realizations of color codes~\cite{Nigg2014, Ryan-Anderson2021, Postler2022, Ryan-Anderson2022, Paetznick2024, Ryan-Anderson2024, Mayer2024, Bluvstein2024, Postler2024} have not addressed performance scaling with code size on any platform.
Here, we present a comprehensive demonstration of the color code on a superconducting processor~\cite{Acharya2024}, achieving logical error suppression and performing logical operations.
Scaling the code distance from three to five suppresses logical errors by a factor of $\errorsuppressionfactor = \lambdaneuralnetworkvalue$. Simulations indicate this performance is below the threshold of the color code, and furthermore that the color code may be more efficient than the surface code with modest device improvements.
Using logical randomized benchmarking~\cite{Combes2017}, we find that transversal Clifford gates add an error of only \errorpercliffordneuralnetworkvalue{}, which is substantially less than the error of an idling error correction cycle.
We inject magic states~\cite{Bravyi2005}, a key resource for universal computation, achieving fidelities exceeding 99\,\% with post-selection (retaining about 75\,\% of the data). 
Finally, we successfully teleport logical states between distance-three color codes using lattice surgery~\cite{Horsman2012}, with teleported state fidelities between \teleportationfidelityneuralnetworkvaluemin{} and \teleportationfidelityneuralnetworkvaluemax. 
This work establishes the color code as a compelling research direction to realize fault-tolerant quantum computation on superconducting processors in the near future.

\end{abstract}

\maketitle
\input{authors_short}

\section{Introduction}
Quantum computing holds immense potential for solving complex problems beyond the reach of classical computers~\cite{aspuru2005simulated, lloyd1996universal, shor1999polynomial}. However, most practical quantum computing applications require gate error rates far below what current physical devices can achieve~\cite{Preskill2018}. 
Quantum error correction (QEC) offers a solution to bridge this gap by encoding logical qubits across multiple physical qubits~\cite{shor1995scheme, gottesman1997stabilizer}. This encoding can exponentially suppress errors, provided the physical error rates are below a critical threshold~\cite{Knill1998}.

Realizing scalable, fault-tolerant quantum computation in practice requires overcoming several core challenges, including achieving low logical error rates with reasonable qubit overhead and performing logical operations efficiently. 
Considerable experimental progress has been made in recent years to address these challenges. 
Notably, scalable error suppression was recently demonstrated by increasing the size of a logical qubit encoded in the surface code~\cite{Acharya2023, Acharya2024} using superconducting circuits~\cite{Blais2021}.
QEC implementations with superconducting qubits~\cite{Andersen2020, Marques2022, Krinner2022,  Zhao2022, Sundaresan2023, Ye2023, Hetenyi2024} have predominantly focused on the surface code~\cite{fowler2012surface} or variants thereof~\cite{bonilla2021xzzx, Chamberland2020a}, although alternative approaches such as bosonic encodings are also being explored~\cite{sivak2023real, Putterman2024}. The surface code offers a high error threshold and is compatible with planar, four-nearest-neighbor architectures. However, it demands significant qubit overhead and has limitations for certain logical operations~\cite{Campbell2017}, motivating the investigation of alternative planar codes. 

Color codes~\cite{Bombin2006, Landahl2011} allow for more efficient logical operations and require fewer qubits than the surface code to encode a logical qubit at a fixed code distance~\cite{Bombin2007}.
The smallest instance of the color code, the Steane code~\cite{Steane1996}, has been implemented with trapped ions~\cite{Nigg2014, Ryan-Anderson2021} and neutral atoms~\cite{Bluvstein2024}, demonstrating logical operations and logical circuits~\cite{Postler2022, Ryan-Anderson2022, Paetznick2024, Ryan-Anderson2024, Mayer2024, Bluvstein2024, Postler2024}.
Color codes can perform all Clifford gates~\cite{Gottesman1998} within a single error correction cycle, and support resource-efficient magic state injection protocols required to implement non-Clifford gates~\cite{Bravyi2005, Zhang2024, Lee2024}. A recent breakthrough harnessed these features to drastically reduce the overhead required for non-Clifford gate implementation in surface code architectures~\cite{Itogawa2024, Gidney2024}. Color codes also enable multi-qubit entangling operations with up to three times less space-time overhead thanks to their capability for simultaneous fault-tolerant multi-qubit Pauli measurements~\cite{Thomsen2022}. 

However, color codes have a stricter error threshold than surface codes due to their higher-weight stabilizer measurements~\cite{Andrist2011, Landahl2011}. Additionally, they require more elaborate decoding strategies, and conventional color code syndrome-extraction circuits require higher connectivity than four-nearest-neighbors~\cite{Landahl2011, Takada2024}, which is difficult to realize on superconducting devices. As a result, error suppression by increasing the color code size has not yet been demonstrated on any experimental platform.

Nevertheless, recent advances in superconducting qubit performance~\cite{Acharya2024}, improvements in decoding algorithms~\cite{Baireuther2019, Sahay2022, Gidney2023, Zhang2024, Lee2024}, and optimized error-syndrome extraction circuits~\cite{Gidney2023, Takada2024} now open new possibilities for implementing the color code on existing hardware. 
Here, we demonstrate building blocks of fault-tolerant quantum computation using the color code on a superconducting processor. 
First, we suppress logical errors by increasing code distance. Our simulations extrapolating to larger distances suggest that with modest improvement in physical error rates, color codes might surpass the surface code in qubit efficiency (see~\cref{app:distance_scaling_simulations} for details). 
Next, we characterize single-qubit logical Clifford gates with randomized benchmarking, realizing logical circuits with up to ten Clifford gates interleaved with quantum error correction cycles. 
Furthermore, we inject magic states into the color code with high fidelity and perform lattice surgery~\cite{Horsman2012} to demonstrate multi-logical-qubit operations. 
This comprehensive demonstration establishes the color code as a promising approach for resource-efficient, fault-tolerant quantum computation on superconducting circuits.

\section{Color codes with superconducting qubits}
Two-dimensional (or triangular) color codes~\cite{Bombin2006, Bombin2013}, are a family of topological stabilizer codes~\cite{gottesman1997stabilizer}. They are constructed on trivalent lattices with three-colorable tiles,
meaning that three tiles meet at each vertex of the lattice and the tiles can be colored (red, green and blue in this work) such that no two adjacent tiles share the same color. Data qubits are located at each vertex of the lattice, and each tile has an X-type and a Z-type stabilizer with supports on its vertices. 

\begin{figure}
    \centering
    \includegraphics[width=\columnwidth]{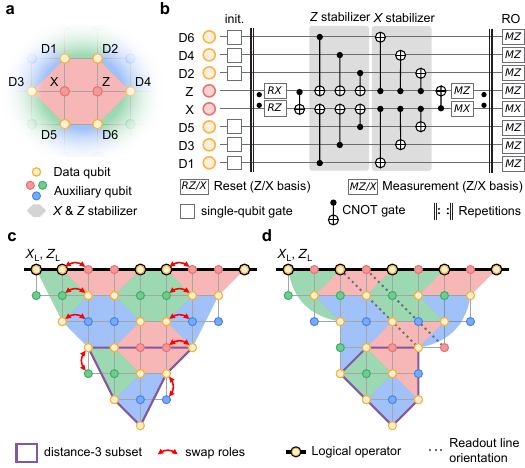}
    \caption{\textbf{Superdense color code.} 
    \textbf{a,} Example tile in the bulk of a red, green, blue hexagonal lattice employed for the superdense color code. The lattice is embedded on a square grid of data qubits (golden circles labeled D1 to D6) and X/Z \auxiliary{} qubits (red, green and blue circles), with their connectivity indicated by solid gray lines. 
    \textbf{b,} Superdense syndrome extraction circuit for the tile shown in \textbf{a}, see text. 
    \textbf{c,} Distance-five color code  qubit, with one of the distance-three qubit subsets outlined in purple. Data qubits included in the logical operators \lopx{} and \lopz{} are circled and connected by a solid black line. Red arrows indicate qubit pairs interchanged for the implementation of the color code on our quantum device. \textbf{d,} Deformed code layout after interchanging the qubits to ensure that each readout line contains only data or \auxiliary{} qubits. The readout lines are oriented diagonally from top left to bottom right, as shown by two dashed gray lines.}
    \label{fig:color_code_concept}
\end{figure}

Several such lattices exist, each with potentially different syndrome extraction circuits for stabilizer measurements~\cite{Gidney2023}. Here, we choose to implement a code based on a hexagonal lattice embedded in a square grid of qubits with a superdense syndrome extraction circuit~\cite{Gidney2023}, shown in \cref{fig:color_code_concept}a-b.
This choice is particularly attractive for superconducting qubits because it only requires nearest-neighbor connectivity on a square grid of qubits. In addition, this choice enables simultaneous measurement of X- and Z-type stabilizers of each tile using two \auxiliary{} qubits  labeled X and Z in \cref{fig:color_code_concept}a, located at the center of each tile. 
Conceptually, the two \auxiliary{} qubits are prepared in a Bell pair and each interact with the three nearest data qubits of the tile using CNOT gates to accumulate first the Z-type, then the X-type stabilizer, as depicted in \cref{fig:color_code_concept}b. A Bell-basis measurement of the \auxiliary{} qubits follows, allowing both stabilizers of the hexagon to be measured simultaneously up to Pauli corrections dealt with by a frame update~\cite{Gidney2023}. 

This construction allows the encoding of a single distance-five logical qubit using 19 data qubits and 18 \auxiliary{} qubits, see \cref{fig:color_code_concept}c.    
It consists of six weight-6 stabilizers (two per hexagonal tile) and twelve boundary weight-4 stabilizers (two per quadrilateral tile).
The logical operator \lopx{} (\lopz{}) is defined as the product of the individual Pauli \px{} (\pz{}) operators of data qubits along a boundary of the triangle. 

We implement this logical qubit on the 72-qubit Willow processor first introduced in Ref.~\onlinecite{Acharya2024}; see \cref{app:device} for a summary of the device performance.
The design of this processor favors the simultaneous readout of all qubits sharing a readout line to minimize measurement-induced dephasing. To accommodate this constraint, we swap the roles of \auxiliary{} and data qubits for certain pairs of qubits, as indicated by the red arrows in \cref{fig:color_code_concept}c. This adjustment results in the code layout depicted in \cref{fig:color_code_concept}d, where each readout line contains only data qubits or \auxiliary{} qubits, but no mixture of both. This ensures that all qubits on each readout line are measured either in each quantum error correction cycle (\auxiliary{} qubit readout line) or at the end of the protocol only (data qubit readout line). To avoid the use of costly swap gate decompositions, we employ circuit transformations that add no additional operations and preserve the circuit’s fault tolerance, as detailed in \cref{app:superdense_circuit}. 
In addition to these circuit transformations, the CNOT gates are compiled to a combination of native Hadamard gates and CZ gates; see \cref{app:circuit_transformation} for a full circuit diagram of a quantum error correction cycle on the distance-five logical qubit.

\section{Distance scaling}
First, we demonstrate that the color code is a viable candidate for encoding logical qubits by testing its memory performance and scaling. 
Specifically, we preserve logical states in the X and Z bases through repeated cycles of error correction, and suppress the logical error per cycle by increasing the code distance $d$ from three to five. Here, the distance of the code corresponds to the minimum number of physical qubit errors required to cause an undetectable logical error. A distance $d$ code can correct any $\lfloor \frac{d-1}{2}\rfloor$ independent errors.
We compare the performance of the distance-five code to the averaged performance of three distance-three color code subsets. An example subset is shown with a purple outline in \cref{fig:color_code_concept}c and d, and a visual representation of all $d=3$ subsets is presented in \cref{app:distance_scaling}.
For each experimental run, we initialize the data qubits in a product state of the Z (X) basis, project the logical qubit into the target logical state using a single cycle of stabilizer measurements, perform $n-1$ additional cycles of error correction and finally measure all data qubits in the Z (X) basis.  We obtain the logical operator value from the product of the relevant data qubits in this final measurement, and the run succeeds if the corrected logical measurement after decoding coincides with the target logical state; otherwise, a logical error has occurred.

We employ similar techniques as described in Refs.~\onlinecite{Acharya2023, Acharya2024} to minimize coherent phase errors in the circuit~\cite{Kelly2016}, reduce dephasing during idle time on data qubits using dynamical decoupling, and ensure short-lived leakage errors with multi-level reset~\cite{mcewen2021removing} on \auxiliary{} qubits and a data-qubit leakage removal (DQLR)~\cite{miao2023overcoming} protocol at the end of each cycle.

\begin{figure}[t]
    \centering
    \includegraphics[width=\columnwidth]{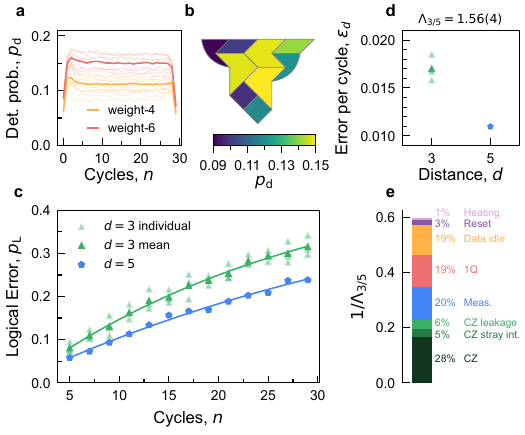}
    \caption{\textbf{Distance scaling experiment.} 
    \textbf{a,} Detection probability \pdet{} as a function of quantum error correction cycle \nmemcycles{} for individual stabilizers (faded lines) and their average (solid line) for an $X$-basis state preservation experiment in a distance-five color code. Weight-6 (weight-4) stabilizers are colored in red (gold). 
    \textbf{b,} Detection probability for each tile of the distance-five color code, averaged over cycles and bases. 
    \textbf{c,} Measured logical error \logicalerrorprob{} for distance-three (green triangles) and distance-five (blue pentagons) codes averaged over the X and Z bases. Faded symbols correspond to individual distance-three subsets. The solid lines, shown for the averaged distance three code and the distance-five code, are fits to $\logicalerrorprob = \varepsilon_0 \cdot (1 - 2 \cdot \epc)^\nmemcycles + 1/2$, with fitting parameters $\varepsilon_0$ and \epc. 
    \textbf{d,} Logical error per cycle, \epc, versus code distance, $d$. Same symbols as \textbf{c}.
    \textbf{e,} Relative contributions of different error sources to the error budget for the color code: CZ errors (CZ), errors from spurious interactions during two-qubit gates (CZ stray int.), leakage errors during two-qubit gates (CZ leakage), measurement errors (Meas.), single-qubit gate error (1Q), data-qubit idle error during measurement and reset of \auxiliary{} qubits (Data idle), reset error (Reset), leakage due to incoherent heating from $\ket{1}$ to $\ket{2}$ (Heating). The relative contributions of the different error channels are indicated.}
    \label{fig:distance_scaling}
\end{figure}

For each error correction cycle, the \auxiliary{} qubit readout outcomes correspond to the stabilizer values and indicate the parity of the involved data qubits (0 for even parity and 1 for odd parity). We construct the error syndrome $\syndrome$, in which each element is obtained by comparing stabilizer values in two consecutive cycles and takes a value of 1 if a change of parity is detected and 0 otherwise. By averaging each syndrome element over 50\,000 experimental runs, we obtain the average error detection probability \pdet{} for each stabilizer in each cycle, see \cref{fig:distance_scaling}a for the error detection probabilities of the distance-five X-basis state preservation experiment \optional{and \cref{app:detection_probabilities} for others}. We find that \pdet{} remains nearly constant in the bulk of the error correction cycles, suggesting a stable error rate throughout the experiment. As expected, weight-6 stabilizers exhibit a higher error detection probability ($\pdet{}_{, 6} = \pdetweightsixdistancefiveX$  when averaging over all cycles and stabilizers), compared to boundary weight-4 stabilizers ($\pdet{} _{, 4} = \pdetweightfourdistancefiveX$), which involve fewer gates and are thus less susceptible to errors, as visualized in \cref{fig:distance_scaling}b.

Because each data qubit in the bulk participates in three stabilizers per basis (compared to two for the surface code), efficient matching-based decoders~\cite{Edmonds1965, Dennis2002, higgott2023sparse, Wu2023} cannot be used directly~\cite{Gidney2023}. Therefore,  we employ novel decoding strategies~\cite{Landahl2011, Gidney2023, Takada2024, Bausch2024} to infer whether logical errors have occurred; see \cref{app:decoding} for an overview of the different decoders used in this work.

To assess the performance of our color code, we initialize the logical qubit, measure up to $\nmemcycles = 29$ cycles of error correction, and compute the logical error probability \logicalerrorprob{} as a function of \nmemcycles. We then fit \logicalerrorprob{} to obtain the logical error per cycle \epc{} ~\cite{Krinner2022, Acharya2023}. We average the error per cycle of the three different distance-three subsets to obtain $\meanepc[3]$ and compute the error suppression factor $\errorsuppressionfactor{} = \meanepc[3] / \epc[5]$. With a neural-network decoder (AlphaQubit)~\cite{Bausch2024}, we obtain $\errorsuppressionfactor{} = \lambdaneuralnetworkvalue$ and $\epc[5] = \epcfivebestvalue$.  

This significant reduction in logical error with increasing distance suggests performance below the error correction threshold of the color code, corroborated by Pauli simulations extrapolating to larger distances and varying noise strengths (see \cref{app:threshold_simulations}).
The observed error suppression factor is in reasonable agreement with the simulated error suppression factor $\errorsuppressionfactorsim = \lambdaneuralnetworksimvalue$. By varying the strength of individual error sources in simulations, we estimate their relative contributions to $1/\errorsuppressionfactorsim$~\cite{Acharya2024}. 
This analysis suggests that CZ gates are the primary contributors, accounting for about 39\,\% of the error budget (including CZ-gate-induced leakage and stray interactions during CZ-gates). The remaining contributions are approximately equally distributed among measurement errors, single-qubit gate errors, and data-qubit idle errors during the measurement and reset of \auxiliary{} qubits.

\section{Logical Randomized Benchmarking}
Achieving fault-tolerant quantum computation requires not only correcting errors in idle logical qubits but also applying logical gates that do not spread errors. 
Transversal single-qubit logical gates~\cite{Eastin2009} achieve this by applying the desired gate independently to each physical data qubit, keeping potential errors isolated. 
They are inherently fault-tolerant, simple to implement, and efficient, as they require only a single time step of physical gates.

\begin{figure}[t]
    \centering
    \includegraphics[width=\columnwidth]{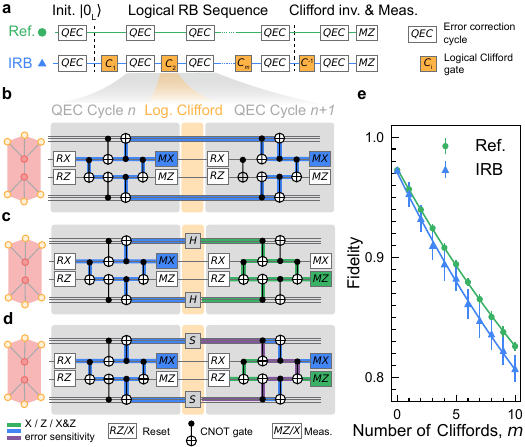}
    \caption{
    \textbf{Logical randomized benchmarking.} 
    \textbf{a,} Logical-qubit-level reference (green) and interleaved randomized benchmarking (blue) circuit diagrams, consisting of error correction cycles (QEC), randomly selected Clifford gates ($C_1, ..., C_\nrclifford$), a Clifford recovery gate  $C^{-1}$ and a Z basis measurement (\textit{MZ}). 
    \textbf{b-d,} Simplified circuit diagrams for a tile of the color code indicating measurements included in an X-stabilizer error-detecting region spanning across two consecutive cycles without logical gate (\textbf{b}), and how it changes upon the application of a logical Hadamard gate $H$ (\textbf{c}), or a logical phase gate $S$ (\textbf{d}) between two error correction cycles, see text for details. 
    The green, blue, and purple highlighted sections correspond to regions where the detecting region is sensitive to Z, X, and both X and Z errors, respectively.
    A controlled-NOT (CNOT) gate symbol spanning three qubit wires indicate three consecutive CNOTs between the \auxiliary{} qubit and its neighboring data qubits.
    \textbf{e,} Measured fidelity (symbols) and exponential fits (solid lines) for the interleaved (blue) and reference (green) sequences versus the number of logical Clifford gates \nrclifford. The data is decoded using the neural-network decoder.  Error bars represent the standard deviation of fidelity over 25 random Clifford sequences, each repeated 20\,000 times.}
    \label{fig:logical_rb}
\end{figure}

One important advantage of the color code over the surface code is its ability to perform all single-qubit logical Clifford gates transversely. 
Here, we implement all 24 gates of the Clifford group using a combination of Hadamard gates $H$, phase gates $S$, and Pauli operators, see \cref{app:logical_rb}. 
We characterize the average error of transversal Clifford operations using logical randomized benchmarking (LRB)~\cite{Combes2017}.
In analogy to interleaved randomized benchmarking (iRB) at the physical-qubit level~\cite{Magesan2012}, the protocol consists of applying multiple sequences of \nrclifford{} randomly chosen logical Clifford gates, each followed by a final Clifford gate $C^{-1}$ that inverts the sequence. The logical qubit is subsequently measured and compared to the initial logical state, \zerological. This process is repeated for various Clifford sequence lengths \nrclifford{}, with a cycle of error correction between each Clifford gate. 
To obtain an estimate of the average error per logical Clifford gate, we compare these sequences to reference sequences in which no Clifford gates are applied i.e. a standard $\zerological$ state-preservation experiment, as depicted in \cref{fig:logical_rb}a.

In a state preservation experiment (no logical gates), subsequent measurements of the same auxiliary qubits are used to construct error syndrome elements. Applying a logical gate between error correction cycles transforms the stabilizers, changing how syndrome elements are constructed, as illustrated in \cref{fig:logical_rb}b-d. For instance, a transversal $H$ gate turns an X stabilizer into a Z stabilizer (and vice-versa). 
Consequently, the X auxiliary qubit measurement in cycle $n$ must be compared to the Z measurement in cycle $n + 1$ to form a valid syndrome element.
Similarly, an $S$-gate transforms X stabilizers into Y stabilizers (leaving Z stabilizers unaffected), such that a phase flip in cycle $n$ flips both X and Z \auxiliary{} measurements in cycle $n+1$. We track stabilizer transformations and determine which measurements to compare in each cycle according to these rules.

We realize LRB in a distance-three color code, varying the number of randomly chosen Clifford gates $\nrclifford=0$ to 10, with 25 random sequences for each \nrclifford{}. 
When compiling these circuits to native physical operations available on our hardware, single-qubit gates implementing the transversal gate are potentially merged with those from adjacent error correction cycles to minimize the total number of operations per qubit. 
We observe an exponentially decaying randomized benchmarking fidelity, with the sequence including transversal Clifford gates decaying slightly faster than the reference idling experiment (\cref{fig:logical_rb}e). 
We extract an average logical-Clifford gate error \errorperclifford, which quantifies the additional error introduced by the logical Clifford gate in each cycle. Using the neural-network decoder, we obtain  $\errorperclifford = \errorpercliffordneuralnetworkvalue{}$. 
This is significantly lower than the logical error per cycle, $\meanepc[3] = \meanepcthreebestvalue$, underscoring the efficiency of implementing transversal single-qubit gates.

\section{Magic State Injection}
Despite their efficient implementation, transversal gates cannot form a  universal gate set~\cite[p.\,188]{Nielsen2010} required for arbitrary error-corrected quantum computation~\cite{Eastin2009}.
For the color code, extending the transversal gate set with a non-Clifford single-qubit $T$-gate ($\pi/4$ rotation around the $z$ axis) suffices to perform all single-qubit logical operations necessary for universal quantum computation. 
This gate can be implemented by preparing a high-fidelity resource magic state~\cite{Bravyi2005} on an auxiliary logical qubit and consuming it in a gate teleportation protocol~\cite{Bravyi2005, terhal2015quantum}.
Preparing the high-fidelity magic state consists of two steps: first, injecting a magic state prepared on a physical qubit into a logical qubit, which is inherently error-prone because it cannot be performed fault-tolerantly like for Pauli eigenstates; and second, distilling very high-fidelity magic states from an ensemble of faulty ones through magic state distillation~\cite{Bravyi2005}. 
Here, we focus on the state injection step.

\begin{figure}[t]
    \centering
    \includegraphics[width=\columnwidth]{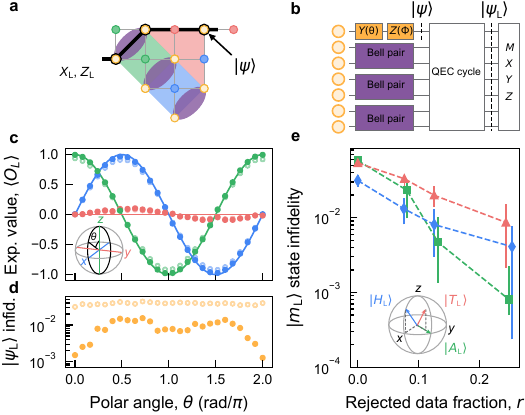}
    \caption{\textbf{Arbitrary state injection in a distance-three color code.} 
    \textbf{a,} Schematic representation of a $d=3$ color code. The arbitrary state $\ket{\psi}$ is prepared on the data qubit indicated with a black arrow. The black line indicates the logical operators of the $d=3$ color code, and the purple ellipses indicate Bell pairs, see text. 
    \textbf{b,} Simplified circuit diagram for the state injection. Each line corresponds to one of the data qubits, on which we apply single-qubit $Y$- and $Z$-rotations (yellow boxes), Bell pair preparation circuits (purple boxes), a quantum error correction (QEC) cycle and a measurement in the X, Y, or Z  basis (\textit{MXYZ}) for logical state tomography. 
    \textbf{c,} Decoded (semi-transparent circles) and post-selected (solid dots) expectation value of the logical Pauli operators \lopx{} (blue), \lopy{} (red), \lopz{} (green) when sweeping the polar angle \polarangle. The solid lines correspond to ideal expectation values. 
    \textbf{d,} Decoded (semi-transparent circles) and post-selected (solid dots) infidelities for the prepared logical state \arbitrarymagicstate. 
    \textbf{e,} Magic state $\ket{m_\mathrm{L}}$ infidelity as a function of rejected data fraction for \magicstateA{} (green squares), \magicstateH{} (blue diamonds), \magicstateT{} (red triangles), see text and \cref{app:state_injection}. The dashed lines serve as a guide to the eye, and the error bars indicate a 95\,\% bootstrapped confidence interval. Each state is depicted on the Bloch sphere by an arrow of the corresponding color.}
    \label{fig:state_injection}
\end{figure}

We begin by preparing an arbitrary state $\ket{\psi}$ on a single data qubit using $Y$- and $Z$-rotations, parameterized by the polar angle \polarangle{} and azimuthal angle \azimuthalangle, respectively. 
We then grow into a distance-three color code by initializing the new data qubits pairwise into Bell states (purple ellipses in \cref{fig:state_injection}a), ensuring the $\hat{X}$ and $\hat{Z}$ operators of the injection qubit (indicated with a black arrow) deterministically extend to \lopx{} and \lopz{} (\cref{app:state_injection})~\cite{Jones2016, Zhang2024}.
Executing a single cycle of error correction thereafter projects the distance-three color code into \arbitrarymagicstate{}; see \cref{fig:state_injection}b for a simplified circuit diagram and \cref{app:state_injection} for details.
Because the protocol is applicable to arbitrary states and starts with a single physical qubit, an error on the injection data qubit cannot be detected.  However, it has been shown to achieve lower error rates than other injection protocols for CSS codes~\cite{Zhang2024}. 

We characterize the injection of arbitrary states by sweeping the polar angle \polarangle{} from 0 to $2\pi$ while keeping $\azimuthalangle = 0$ and performing logical tomography, measuring the expectation values of \lopx{}, \lopy{}, and \lopz. As expected for a rotation around the $y$ axis of the logical Bloch sphere, $\langle\lopz{}\rangle$ and $\langle\lopx{}\rangle$ oscillate with sinusoidal shapes, while $\langle\lopy{}\rangle$ remains close to zero (see \cref{fig:state_injection}c).
From the measured values, we compute the state infidelity across the sweep, obtaining an average infidelity of \polarangleinfidchromobiusvalue{} using the Möbius decoder (Chromobius)~\cite{Gidney2023} and \polarangleinfidpostselectionvalue{} when post-selecting for runs without detectable errors (keeping on average \polaranglepostselectionfraction{} of the runs), see \cref{fig:state_injection}d. Note that post-selection is acceptable in this context because state injection protocols can be repeated until a high-fidelity resource state is achieved with high confidence~\cite{Li2015, Litinski2019, Lao2022}.

We perform a similar analysis for the azimuthal angle sweep (detailed in \cref{app:state_injection}) and then focus on preparing specific magic states. We examine $\magicstateA = \frac{1}{\sqrt2}(\zerological + e^{\i\pi/4}\onelogical)$, $\magicstateH = \cos\frac{\pi}{8}\zerological + \sin\frac{\pi}{8}\onelogical$, and $\magicstateT = \cos\frac{\beta}{2}\zerological + e^{\i\pi/4}\sin\frac{\beta}{2}\onelogical$ (with $\beta = \arccos 1/\sqrt{3}$), corresponding to the +1 eigenstate of the $\lopx+\lopy$, $\lopx + \lopz$,  and $\lopx + \lopy +\lopz$ operators, respectively.
We achieve post-selection infidelities of \magicstateAinfidpostselectionvalue, \magicstateHinfidpostselectionvalue, and \magicstateTinfidpostselectionvalue{} with retained data fractions of \magicstateAfullpostselectionfraction, \magicstateHfullpostselectionfraction, and \magicstateTfullpostselectionfraction, respectively. Here, the uncertainty represents a 95\,\% confidence interval calculated using bootstrapping. 
We also explore partial post-selection, balancing data rejection with infidelity (see \cref{fig:state_injection}e). These high-fidelity magic states surpass the threshold required for magic state distillation~\cite{Bravyi2005, Reichardt2005}, exceed the fidelities achieved in other state injection experiments in superconducting circuits~\cite{Marques2022, Ye2023, Gupta2024, Kim2024}, and lie on the Pareto front balancing infidelity and data rejection when compared to state injections on other platforms (see \cref{app:state_injection}).

\section{State teleportation using lattice surgery}
Lattice surgery~\cite{Horsman2012, Landahl2014, Thomsen2022, Zhang2024} enables the efficient realization of multi-qubit fault-tolerant operations, such as the CNOT-gate, with only nearest-neighbor interactions. 
In essence, it performs these operations using fault-tolerant measurements of multi-qubit logical Pauli operators, combined with conditional operations based on the measurement outcomes. 

\begin{figure}[t]
    \centering
    \includegraphics[width=\columnwidth]{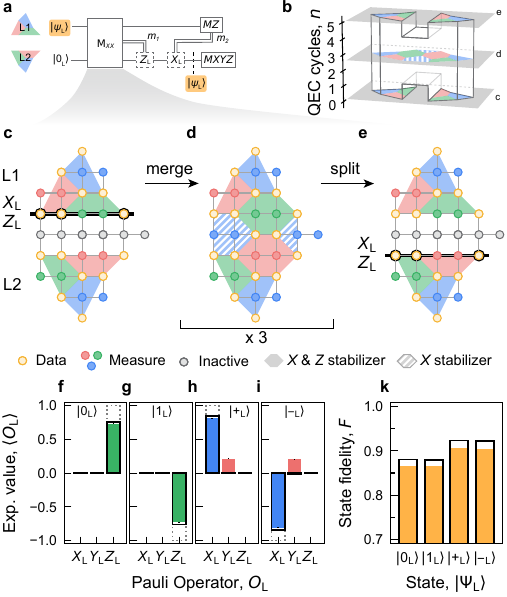}
    \caption{\textbf{State teleportation in the color code using lattice surgery.} 
    \textbf{a,} Simplified circuit diagram to teleport a state \teleportationstate{} from logical qubit L1 to a logical qubit L2 using an $XX$-parity measurement (\mxx) realized by lattice surgery, and Pauli frame updates conditioned on the measurement outcomes \teleportationmxxoutcome{} and \teleportationmzoutcome{} performed in post-processing (dashed boxes). 
    \textbf{b,} Space-time block diagram illustrating the \mxx{} lattice surgery operation, with horizontal cuts displaying the evolution of the stabilizers as a function of the quantum error correction (QEC) cycles, as detailed in \textbf{c}-\textbf{e}. 
    \textbf{c-e,} Representation of the active qubits (colored) and stabilizers before (\textbf{c}) and after (\textbf{d}) the merge operation, and after the split operation (\textbf{e}). The hatched tiles indicate stabilizers in the X basis only, while filled tiles indicate both X and Z stabilizers with support on its vertices. The black line indicates the teleported logical operator from L1 to L2, after the Pauli frame update.
    \textbf{f-i,} Measured (color), simulated (black wireframe), and ideal (dotted gray wireframe) expectation values of the Pauli logical operators of L2 after teleporting state \zerological, \onelogical,\pluslogical, and \minuslogical, respectively.
    \textbf{k,} Measured (gold bars) and simulated (black wireframe) teleported state fidelity $F$ of the four logical X and Z eigenstates.
    }
    \label{fig:state_teleportation}
\end{figure}

Here, we showcase the lattice surgery framework using an \mxx{} parity measurement, which measures the two-qubit Pauli operator \lopxx, along with a classical Pauli frame update to teleport~\cite{Bennett1993, Ryan-Anderson2024} a logical state, \teleportationstate, from one logical qubit to another, represented schematically in \cref{fig:state_teleportation}a.
The protocol starts with two distinct distance-three color code logical qubits, L1 and L2, initialized in \teleportationstate{} and \zerological, respectively, using a single QEC cycle. 
We then perform the \mxx{} parity measurement, which involves a three-cycle lattice surgery merge operation, followed by a split operation over one cycle, see \cref{fig:state_teleportation}b for a space-time visualization of the code evolution and \cref{fig:state_teleportation}c-e for spatial cuts at the key stages.
 In the first merge cycle, two new X-basis stabilizers are added, indicated by the hatched tiles in \cref{fig:state_teleportation}d. In addition, the green (red) boundary stabilizers of L1 (L2) are expanded from weight-four to weight-six stabilizers in both bases, requiring the addition of two data qubits initially prepared in a Bell state. 
At the end of the third merge cycle, these data qubits are measured in the Bell basis, and the parity measurement outcome \teleportationmxxoutcome{} is determined from the product of the X-basis merge stabilizers and the Bell measurement outcome.
In the subsequent QEC cycle, we stop measuring the merge stabilizers and perform instead a QEC cycle on each patch in its original configuration to complete the split operation.
Finally,  L1 is measured in the Z basis while we perform logical state tomography on L2 to reconstruct the teleported state density matrix. 
The parity measurement outcome \teleportationmxxoutcome{} and L1's Z-basis measurement outcome \teleportationmzoutcome{} are used in post-processing to apply a conditional Pauli frame update $\lopz^{\teleportationmxxoutcome} \lopx^{\teleportationmzoutcome}$ to the state of L2, effectively completing the teleportation of the logical operators indicated by the black line in \cref{fig:state_teleportation}c from L1 to L2.
We provide a more detailed explanation of the protocol along with an example circuit diagram in \cref{app:state_teleportation}.

We experimentally realize the teleportation protocol for four input states (\zerological, \onelogical, \pluslogical{} and \minuslogical), and measure the expectation value of the logical Pauli operators \lopx, \lopy, \lopz{} of the output state. 
For each state, only the expectation value in its corresponding eigenbasis (Z basis for \zerological{} and \onelogical, X basis for \pluslogical{} and \minuslogical) shows a large deviation from zero after decoding, see \cref{fig:state_teleportation}f-i. 
The decoded eigenvalues are in general agreement with the ideal expectations (dotted gray wireframe), and reasonably match Pauli noise simulations based on device benchmarks (black wireframe), except for $\langle\lopy\rangle$ in the X basis states which we discuss in \cref{app:state_teleportation_bias}. Calculating fidelity with the ideal logical states yields fidelities between \teleportationfidelityneuralnetworkvaluemin{} and \teleportationfidelityneuralnetworkvaluemax{} when decoded with the neural-network decoder (\cref{fig:state_teleportation}k). 
These measurements enable us to bound the average fidelity of the single-qubit teleportation channel, \teleportationchannel, at $\teleportationaveragefidelity(\teleportationchannel)\geq\teleportationentanglementfidelityneuralnetworkvalue$ (see \cref{app:state_teleportation} for details).

\section{Discussion and outlook}
In this work, we demonstrate key building blocks of fault-tolerant quantum computation using the color code on a superconducting processor. Combining improved device performance, compact syndrome extraction circuits, and novel decoding strategies, we achieve clear error suppression with $\errorsuppressionfactor{} = \lambdabest$ when scaling the code from distance-three to distance-five. We characterize transversal single-qubit logical Clifford gates using randomized benchmarking. Additionally, we demonstrate high-fidelity magic state injection and successfully teleport logical Pauli states between distance-three qubits using lattice surgery.

These results highlight that quantum error correction with superconducting qubits has reached a pivotal stage where the noise threshold may no longer be the main driver of code choice, and efficient logical operations are a crucial factor. At present, the surface code achieves better logical error suppression (on the same processor, $\Lambda^\mathrm{sc}_{3/5}=2.31(2)$~\cite{Acharya2024}), but simulations suggest that improving physical error rates by about a factor of four could render the color code more qubit-efficient than the surface code (\cref{app:distance_scaling_simulations}). Importantly, the color code’s efficient transversal Clifford gates and more flexible lattice surgery would simplify algorithm implementations~\cite{Thomsen2022}. Color codes also underpin the most efficient methods to generate high-fidelity magic states~\cite{Itogawa2024, Gidney2024}, which could be used alongside surface code computation~\cite{Litinski2017, PoulsenNautrup2017, Shutty2022}. Quantifying the practical impact of these advantages for large-scale algorithms requires detailed resource estimation studies.

Key goals for future research with color codes include increasing device performance and size, improving decoding speed and accuracy, demonstrating logical gate error suppression with growing code size, and characterizing more of the color code's unique capabilities for logical gates.
As hardware continues to improve, other quantum error-correcting codes may also become available for practical implementation. Together, we expect these developments to accelerate progress toward scalable and resource-efficient fault-tolerant quantum computation.

\section*{Acknowledgments}
Nathan Lacroix is thankful to Andreas Wallraff for enabling the research stay at Google Quantum AI. 
The Google DeepMind team would like to thank George Holland, Charlie Beattie, and Toby Sargeant for their support.
We thank the broader Google Quantum AI team for supporting and enabling this work.

\section*{Contributions}
\Nathan, \AlexandreB, \Craig{}, and \KevinS{} designed the experiments with inputs from \Matt{} and \Cody.  
\Nathan{} and  \AlexandreB{} calibrated the device and performed the experiments with the support of \AndreasB, \Jimmy, \Matt{}, \Alexis, and \KevinS.  
\Nathan{} analyzed the data with support from \AlexandreB. 
\Nathan, \Johannes, \Francisco, \Lei, \AlexandreB{}, and  \Noah{} decoded the data with support from \Vlad{} and \Oscar. \Adam{} provided input regarding average fidelity of the teleportation channel.
\Noah, \Oscar, \Nathan, \Jahan, and \Dvir{}  performed the numerical simulations. 
\Nathan{} wrote the manuscript, with input from all co-authors. 
\Julian, \Cody, \Craig{}, and \KevinS{} supervised the work.
The Google DeepMind and Google Quantum AI teams jointly developed the machine learning decoder used.
The Google Quantum AI team sponsored the project, designed and fabricated the device, built and maintained the cryogenic and control systems, provided software infrastructure, performed the initial bring up of the device, and provided relevant guidance.

\section*{Competing interests}
The authors declare no competing interests.

\section*{Supplementary information}
Supplementary Information is available for this paper.
Correspondence and requests for materials should be addressed to A.~Bourassa (abourassa@google.com), N.~Lacroix (nathan.lacroix@phys.ethz.ch), and K.~J.~Satzinger (ksatz@google.com).

\section*{Data availability}
Data is available from the corresponding author upon reasonable request, or at \href{https://doi.org/10.5281/zenodo.14238944}{https://doi.org/10.5281/zenodo.14238944}.

\bibliography{resources/references}

\clearpage
\newpage
\newcommand{\partofmain}{}

\end{document}


\fi

\title{\puttitle}
\title{Supplemental Information for: \puttitle}

\author{Google Quantum AI and Collaborators}
\date{December 18, 2024}

\maketitle

\appendix
\tableofcontents

\section{Comparison of Qubit Resources Estimates for the Superdense Color Code and the Surface Code}\label{app:distance_scaling_simulations}

Estimating the number of physical qubits required to reach algorithmically relevant error rates ($<10^{-6}$) on a single logical qubit is critical for evaluating the resources required for large-scale quantum computation.
The superdense color code requires $n_{\mathrm{cc}}=\frac{3}{2}d^2-\frac{1}{2}$, to encode a distance $d$ logical qubit where $n_{\mathrm{cc}}$ is the number of qubits in the circuit (including \auxiliary{} qubits).
This improves on the rotated surface code circuit, which requires $n_{\mathrm{sc}}=2d^2-1$ qubits -- about 25\,\% more at a fixed distance. 

However, the codes have different noise thresholds and their error suppression factor depend on the physical error rates of the device.
Therefore, to assess the qubit overhead needed to reach algorithmically relevant error rates for each code, we compare them under realistic noise conditions of near-term processors.
Specifically, we simulate both codes using an SI1000 error model~\cite{Gidney2021a} with noise parameters $p=10^{-3}$ (about two times smaller than the error rates of our current device) and $p=5 \cdot 10^{-4}$, for increasing code distances. We decode the resulting data with the MLE decoder to determine the logical error per round for each code and distance. Here, a round is defined as $d$ cycles of error correction.

\begin{figure}[t]
    \centering
    \includegraphics[width=\linewidth]{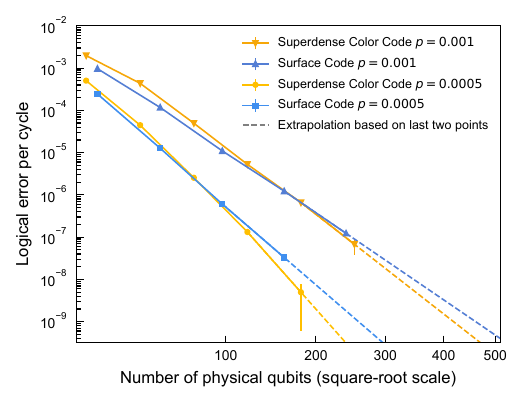}
    \caption{\textbf{Numerical simulations of the superdense color code and surface code.} 
    Simulated logical error per cycle for a qubit encoded in the superdense color code as in Ref.~\onlinecite{Gidney2023} (yellow) and the surface code (blue) for increasing code distances. The error per cycle is plotted as a function of the number of physical qubits required to encode the corresponding logical qubit. Simulations are based on samples taken from an SI1000 error model~\cite{Gidney2021a} with noise parameter $p=10^{-3}$ and $p=5 \cdot 10^{-4}$ decoded using a most-likely-error decoder implemented as a mixed integer program (\cref{app:decoding}). The error bars are calculated from a 90\% confidence interval of a binomial test. Dashed lines correspond to linear extrapolations based on the logical error per cycle of the highest-two code distances. }
    \label{fig:supplement_color_code_vs_surface_code_scaling}
\end{figure}

As expected, the logical error per round decreases with increasing code distance (and thus the number of qubits used), see \cref{fig:supplement_color_code_vs_surface_code_scaling}.
The results indicate that the surface code and the superdense color code will require a comparable number of qubits to achieve a given logical error per round. At large distances, the simulations suggest that the color code may require slightly fewer qubits than the surface code to reach logical error rates on the order of $\sim10^{-8}$, particularly at $p=5 \cdot 10^{-4}$.

This finding is promising, as it suggests that the two codes require similar resources at the physical error rates expected in near-term processors. This allows the choice of error-correcting code to be guided by other considerations, such as the ability to perform logical gates efficiently -- a distinct advantage demonstrated in this work for the color code. 
On a longer horizon where physical error rates get even smaller, the color code is expected to eventually outperform the surface code in terms of qubit overhead.

We observe that the logical error per round for the superdense color code does not follow a linear trend on the square-root scale. Namely, the error suppression factor does not remain constant but increases with code distance. We attribute this effect to spacial boundary effects of triangular color code at small $d$, which become less relevant at larger distances. 

\section{Device and Experimental Setup} \label{app:device}
We implement our experiments on a 72-qubit device first introduced in Ref.~\onlinecite{Acharya2024}.
The device is installed at the bottom of a dilution refrigerator~\cite{Krinner2019} and wired to custom room-temperature electronics for control. 
We use a single control line per flux-tunable transmon~\cite{koch2007charge}
qubit to realize single-qubit gates and to tune the transition frequency to implement two-qubit gates and multi-level reset~\cite{mcewen2021removing}.
We implement single-qubit gates using microwave DRAG pulses~\cite{Motzoi2009} for $X$ rotations and virtual $Z$ gates~\cite{McKay2017}. $\pi$-rotations and $\pi/2$-rotations have durations of \ns{35} and \ns{18}, respectively.
Controlled-$Z$ (CZ) gates between neighboring qubits are realized by tuning the transition frequencies of the two qubits such that the $\ket{11}$ state is resonant with the $\ket{20}$ state, while also adjusting the frequency of a tunable coupler to control the coupling strength. These CZ gates have a duration of \ns{37}.

Qubit states are read out dispersively by applying frequency-multiplexed microwave pulses. The readout protocol,  including cavity depletion time, lasts for \ns{600}. Following readout, we perform a 160-ns multi-level reset as described in Ref.~\onlinecite{mcewen2021removing}. For state preservation experiments, we also include a data qubit leakage removal (DQLR) procedure lasting \ns{53}.

\begin{figure}
    \centering
    \includegraphics[width=\linewidth]{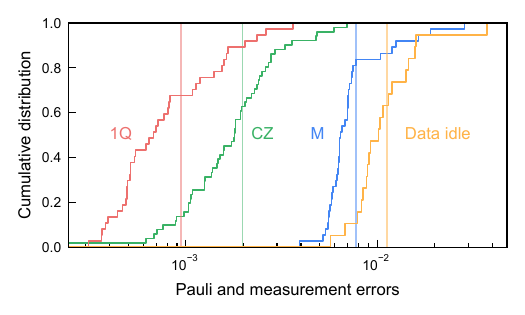}
    \caption{\textbf{Operation errors of the distance-five color code on the 72-qubit processor.} Cumulative distributions of single-qubit gate Pauli errors (red), simultaneous CZ gate Pauli errors (green),  measurement errors (blue), and data-qubit idle errors during measurement and reset of \auxiliary{} qubits when applying a dynamical decoupling sequence (gold). The vertical faded lines indicate the mean value of the distributions.}
    \label{fig:device_performance}
\end{figure}

\begin{figure*}
  \centering
  \includegraphics{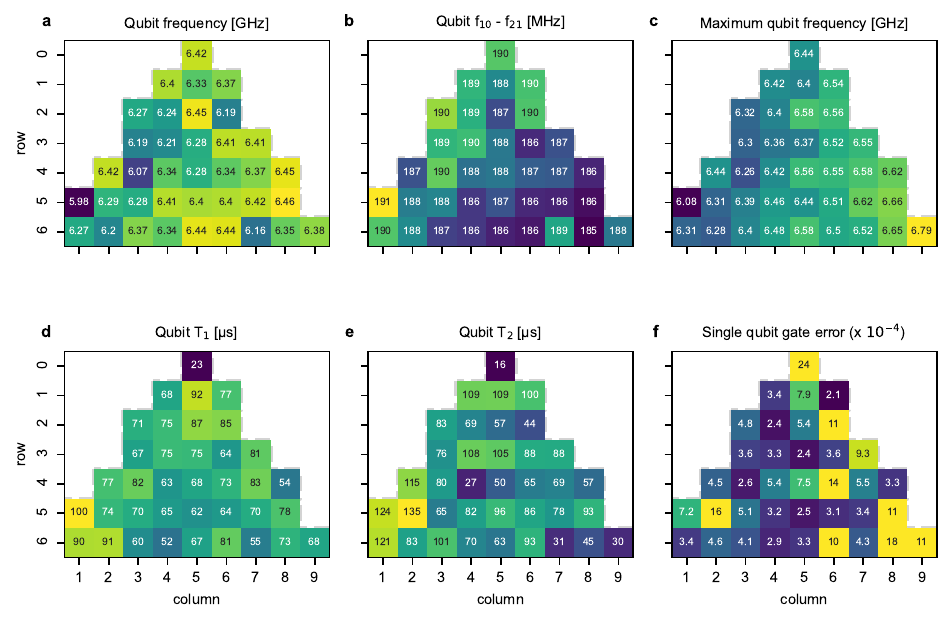}
  \caption{\textbf{Single-qubit parameters.}
      \textbf{a,} Qubit operating frequency during idling and microwave gates.
      \textbf{b,} Qubit anharmonicity.
      \textbf{c,} Maximum qubit frequency.
      \textbf{d,} Qubit relaxation at the operating frequency.
      \textbf{e,} Qubit dephasing at the operating frequency as measured with CPMG echoing ($T_{2,\text{CPMG}}$).
      \textbf{f,} Simultaneous single-qubit gate error as measured by randomized benchmarking (``average'' error).
  }
  \label{fig:qubit_specs}
\end{figure*}

\begin{figure*}
  \centering
  \includegraphics{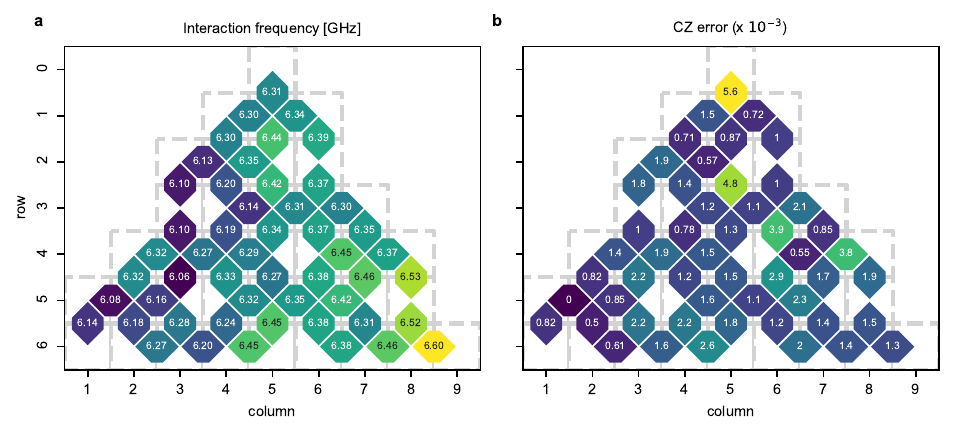}
  \caption{\textbf{Two-qubit gate parameters.}
      \textbf{a,} CZ interaction frequency as defined by the average of the two qubits' $f_{10}$ transitions in the middle of the gate.
      \textbf{b,} Simultaneous CZ XEB gate error. Gate errors are measured in the context of applying all gates in a surface code gate layer. Inferred ``average'' CZ error after subtracting single-qubit error.
  }
  \label{fig:2q_specs}
\end{figure*}

\begin{figure*}
  \centering
  \includegraphics{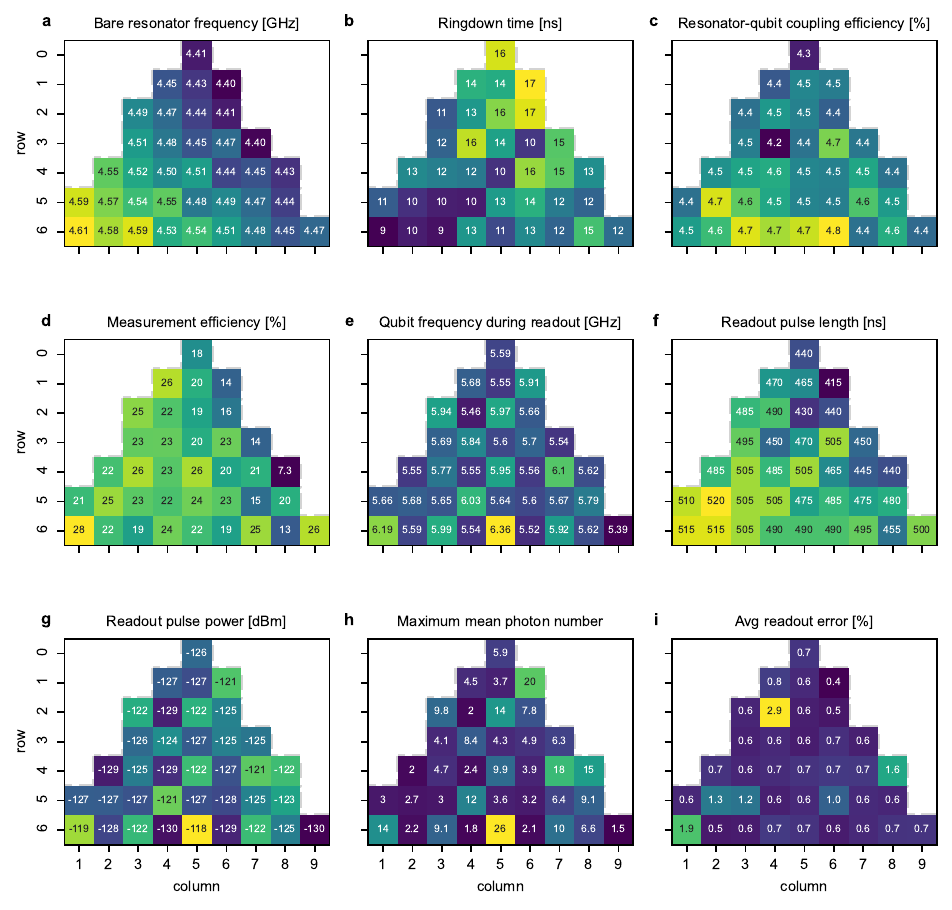}
  \caption{\textbf{Readout parameters.}
      \textbf{a,} Bare frequency of each readout resonator.
      \textbf{b,} Ringdown time (equivalent to $1/\kappa$ of each readout resonator).
      \textbf{c,} The coupling efficiency ($k_{qr}$) between resonator and qubit. The qubit-resonator coupling strength is given by $g_{qr}=k_{qr}\sqrt{\omega_q \omega_r}/2$, where $\omega_q$ and $\omega_r$ are the qubit and resonator frequencies respectively.
      \textbf{d,} The measurement efficiency of the readout. This is the fraction of available information of the heterodyne measurement available after signal amplification and digitization.
      \textbf{e,} Qubit frequency during readout (in the absence of a readout drive).
      \textbf{f,} Length of the applied readout tone on each readout resonator.
      \textbf{g,} Applied power to each readout resonator. 
      \textbf{h,} Simulated maximum mean photon number in each readout resonator, as calibrated using the AC-Stark shift.
      \textbf{i,} Average readout error, $(P(0|1) + P(1|0))/2$, when preparing random states on all qubits.
  }
  \label{fig:ro_specs}
\end{figure*}

We benchmark all operations of the distance-five color code and show the cumulative distribution of error rates in \cref{fig:device_performance}.
We display device maps of single-qubit parameters in \cref{fig:qubit_specs}, two-qubit CZ gate parameters in \cref{fig:2q_specs}, and readout parameters in \cref{fig:ro_specs}. The coordinate system used for these maps matches the one depicting the code instances locations on the processor, see  \cref{app:color_code_location_on_processor} for details.

\section{Superdense Syndrome Extraction Circuit} \label{app:superdense_circuit}

\subsection{Circuit Intuition}
The superdense syndrome extraction circuit for the color code, introduced in Ref.~\onlinecite{Gidney2023}, draws inspiration from the superdense coding protocol, where a Bell pair enables the encoding of two classical bits~\cite{Bennett1992}. 
The superdense circuit is constructed such that these two classical bits indicate whether a bit-flip error occurred (encoded as 0 for no error, 1 for an error) and/or a phase-flip error occurred (encoded as 0 for no error, 1 for an error) on the data qubits of a given color code tile.

The intuition behind this is that the two auxiliary qubits of the tile are prepared in a Bell state, i.e.~a shared quantum resource. This Bell state encodes bit-flip errors in its parity, toggling between even and odd parity depending on whether a bit-flip error occurred on the neighboring data qubits. Simultaneously, it encodes phase-flip errors in its relative phase, which flips between positive and negative depending on whether a phase-flip error occurred. This simultaneous encoding allows the circuit to identify both types of errors using a single round of measurements.

The details of this process can be understood by analyzing how the Bell state evolves throughout the syndrome extraction circuit. Initially, the auxiliary qubits are prepared in the Bell state $\ket{B_{00}} = \ket{00} + \ket{11}$, where the first index (0) represents even parity (symmetric state), and the second index (0) indicates a positive relative phase. Due to this symmetry, a bit flip on either auxiliary qubit transforms the state to $\ket{B_{10}} = \ket{01} + \ket{10}$, with odd parity (antisymmetric state). This property allows each auxiliary qubit to check for bit-flip errors on its three neighboring data qubits using CNOT gates, where the data qubits act as control. Each CNOT toggles the Bell state between $\ket{B_{00}}$ and $\ket{B_{10}}$ if the data qubit is in the $\ket{1}$ state. After all six CNOT gates (left gray box in \cref{fig:color_code_concept}b of the main text), the Bell state will remain $\ket{B_{00}}$ if the total number of flips is even, and flip to $\ket{B_{10}}$ otherwise.

Similarly, a phase flip on either auxiliary qubit changes the state to $\ket{B_{01}} = \ket{00} - \ket{11}$, where the second index (1) indicates a negative relative phase. This enables detection of phase flips on the data qubits by using the auxiliary qubits as control in CNOT gates (right gray box in \cref{fig:color_code_concept}b of the main text). If both bit and phase flips occur, the state becomes $\ket{B_{11}} = \ket{01} - \ket{10}$.

Finally, a Bell-basis measurement of the auxiliary qubits determines the resulting Bell state, producing outcomes $00,\ 01,\ 10$ or $11$ corresponding to $\ket{B_{01}},\ \ket{B_{01}},\ \ket{B_{01}}$ and $\ket{B_{11}}$, respectively. Here, the first bit is the outcome of the Z auxiliary qubit (detects bit-flip errors) and the second bit is the outcome of the X auxiliary qubit (detects phase-flip errors). 
This process reveals whether a bit-flip error, a phase-flip error, both, or none occurred on the data qubits. Remarkably, this circuit achieves this even though each auxiliary qubit interacts with only half the data qubits of the tile.

Note that the Z auxiliary qubit also detects bit-flip errors on the X auxiliary qubit and vice-versa. In that way, the two auxiliary qubits act as flags for each other -- an important feature for ensuring the fault-tolerance of the circuit, which is analyzed in greater detail in the following subsection.

\subsection{Fault-tolerance of the Superdense Syndrome Extraction Circuit}\label{app:fault_tolerance_superdense}
It is generally not guaranteed that a syndrome extraction circuit for a given code preserves the full code distance. For instance, the conventional syndrome extraction circuit for the color code~\cite{Gutierrez2019, Bermudez2019, Chamberland2020, Takada2024} requires flag qubits~\cite{Chao2018} to ensure fault-tolerance.
In this Appendix we provide evidence that the superdense syndrome extraction circuit~\cite{Gidney2023} for the color code preserves the full distance of the color code.

We define the circuit distance \circuitdistance{} to be the minimum number of independent error mechanisms in the syndrome extraction circuit that must occur to flip the logical operator without causing any syndrome element to flip.
In general, $\circuitdistance{}\leq d$, where $d$ is the distance of the code (the minimum weight of a logical operator in the code).
It is well known that the standard surface code syndrome extraction circuit preserves the code distance ($\circuitdistance{} = d$), provided that the sequence of gates is chosen correctly to avoid hook errors~\cite{fowler2012surface}.
However, the circuit distance \circuitdistance{} of the superdense color code circuits from Ref.~\onlinecite{Gidney2023} has not previously been studied.

We verify the circuit distance of the superdense color code using two different methods. The first one is an exact method that consists in mapping the task of determining the code distance to a Maximum Satisfiability (MAxSAT) problem~\cite{Krentel1988}. 
MaxSAT is an optimization problem where the goal is to find the maximum number of constraints that can be satisfied simultaneously in a given Boolean formula. This approach works for our problem because determining the code distance involves finding the smallest set of independent errors that lead to an undetectable logical error, which can be encoded as a set of Boolean constraints. 
To generate the set of constraints for our circuits, we use the stim software package to create a circuit-level noisy representation of the superdense syndrome extraction circuit, where each operation is susceptible to potential errors. We then employ \verb|stim.Circuit.shortest_error_sat_problem()| to produce the corresponding set of constraints in the WDIMACS format.
These constraints are solved exactly as a MaxSAT problem using the \verb|CASHWMaxSAT-CorePlus| MaxSAT solver, which provides the exact circuit distance. Using this method, we compute the circuit distance for distances 3, 5, and 7, confirming that $\circuitdistance = d$.

Solving the MaxSAT problem to determine the circuit distance is NP-hard~\cite{Krentel1988}, which makes it increasingly challenging to compute the exact distance for larger circuits.  Therefore, we employ a heuristic method to search for undetectable logical errors in circuits with larger distances (up to $d=15$). Specifically, we use the \verb|Circuit.search_for_undetectable_logical_errors| function in stim (with the first two arguments set to 3).
Note that this heuristic approach only provides an upper bound on the circuit distance. Nonetheless, the results also suggest that the superdense circuit preserves the full code distance for all distances tested.

\subsection{Superdense Circuit Transformations}\label{app:circuit_transformation}
\begin{figure*}
    \centering
    \includegraphics{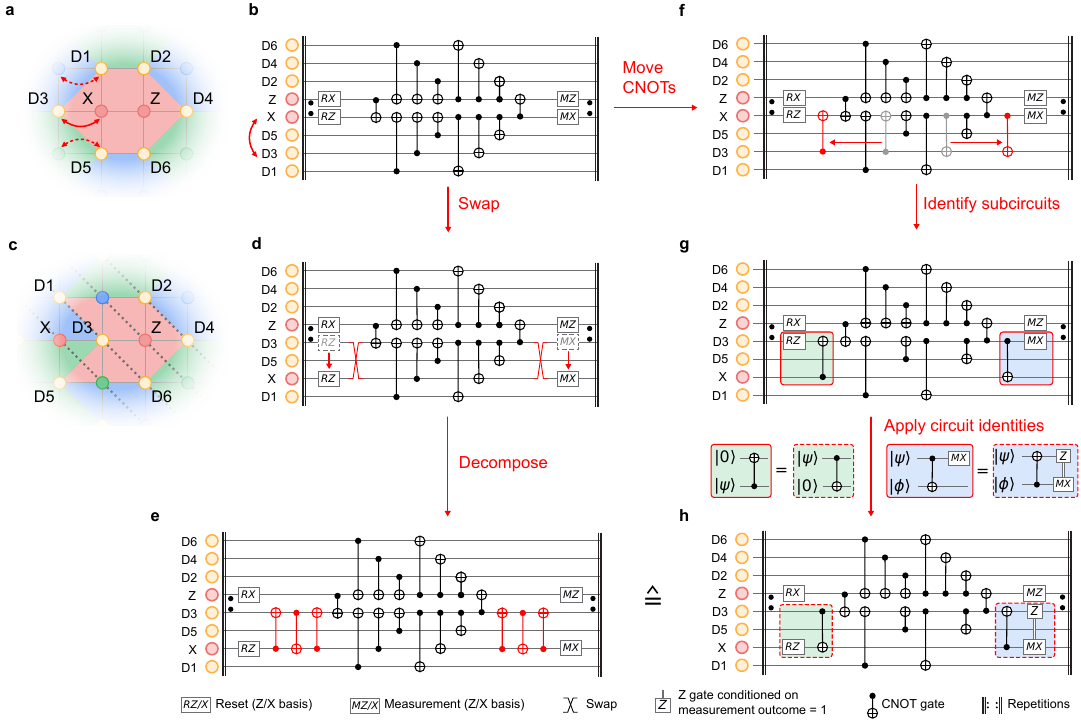}
    \caption{\textbf{Transformation of the superdense syndrome extraction circuit.} Schematic representation of a hexagonal tile in the bulk of a color code (\textbf{a}) and its corresponding syndrome extraction circuit (\textbf{b}). The solid red arrow indicates the pair of qubits of this tile that will be swapped in our implementation. The dashed red arrows indicate similar swaps performed on neighboring tiles but whose circuit transformations are not shown in \textbf{b}. Schematic representation of a deformed tile (\textbf{c}) and its corresponding syndrome extraction circuit (\textbf{d}) with swap gates before the measurement and after the reset. \textbf{e} Syndrome extraction circuit for a naive swap gate decomposition. The CNOT highlighted in red correspond to the swap gates depicted in \textbf{d}. \textbf{f-h} Circuit transformations leading to a logically equivalent circuit as in \textbf{e} but with less physical operations (see text for details).}
    \label{fig:circuit_transformation}
\end{figure*}

Here, we detail how our implementation of the superdense syndrome extraction circuit differs from the original one introduced in in Ref.~\onlinecite{Gidney2023}.
The circuit shown in \cref{fig:color_code_concept}b of the main text (reproduced in \cref{fig:circuit_transformation}b for convenience) is identical to the one shown in Ref.~\onlinecite{Gidney2023}, except for two modifications: we do not show the conditional Pauli flips on data qubits that are handled in pre-processing~\cite{Gidney2023}, and we invert the orientation of the final CNOT along with the corresponding measurement bases of the two \auxiliary{} qubits, for reasons that will become evident below.
This is possible because the two orientations are nominally equivalent sub-circuits for performing Bell measurements. 

When a qubit is measured, the readout pulse can unintentionally dephase other qubits that are not being measured~\cite{Heinsoo2018}. On our device, this effect is more pronounced for qubits located on the same readout line as the measured qubit. To execute this circuit on our processor with best performance, we therefore apply additional transformations to ensure that all qubits sharing a readout line are measured simultaneously. Specifically, we swap the roles of \auxiliary{} and data qubits for specific pairs, as indicated by the red arrows in \cref{fig:color_code_concept}c of the main text, such that each readout line contains only \auxiliary{} qubits or only data qubits but no mixture of both types. 

This transformation is illustrated for a single tile in \cref{fig:circuit_transformation}.
The most straightforward approach involves moving the measurement and reset operations from the X \auxiliary{} qubit to its lateral data qubit neighbor, D3, and introducing swap gates between these qubits before the measurement and after the reset in each cycle (see \cref{fig:circuit_transformation}d). Effectively, this swaps the role of the  X \auxiliary{} qubit and qubit D3 (we therefore relabel the qubits accordingly in \cref{fig:circuit_transformation}d).

Since our device does not support direct implementation of swap gates, each swap must be decomposed into three two-qubit gates (\cref{fig:circuit_transformation}e), which adds six layers of two-qubit gates to the error correction cycle. 
To circumvent this overhead, we implement a logically equivalent circuit (\cref{fig:circuit_transformation}h) that increases the cycle by only two  layers of two-qubit gates by reusing operations already present in the cycle. 
This optimized circuit is derived from the one shown in \cref{fig:circuit_transformation}b by commuting  CNOT gates between the X \auxiliary{} qubit and data qubit D3 (\cref{fig:circuit_transformation}f) outwards and applying circuit identities, as highlighted in the green and blue boxes in \cref{fig:circuit_transformation}g-h.
The final circuit contains a $Z$ gate applied to D3 conditioned on the measurement outcome of the X \auxiliary{} qubit being 1. Similarly to other conditional Pauli operations in this circuit, this gate is not applied physically but instead commuted through the circuit in pre-processing by our error correction software~\cite{Gidney2021}.

Applying this circuit transformation to every tile of the code results in a deformed code layout (\cref{fig:circuit_transformation}c) where each readout line (dotted gray line) exclusively contains either data qubits or \auxiliary{} qubits. To achieve this using the described transformations— where the X \auxiliary{} qubit is swapped with its lateral data qubit neighbor — neighboring tiles must have a mirrored layout for the X and Z \auxiliary{} qubits. At the code boundaries, where the X \auxiliary{} qubit has no lateral neighbor, the swap is performed with the upper data qubit neighbor (D1) instead, as shown in \cref{fig:color_code_concept}c of the main text. Note that these circuit transformations are applicable to arbitrary distances of the superdense color code. 

Finally, before executing the circuit on our hardware, the CNOT gates are decomposed into a combination of Hadamard and CZ gates which are natively supported by our processor. The compiling software optimizes this decomposition by canceling consecutive Hadamard gates on the same qubit to minimize the number of gates applied to each qubit, see also \cref{app:distance_scaling}. 
The total error correction cycle duration after decomposition and optimization is approximately 1.3\,µs.

\begin{figure*}
    \centering
    \includegraphics[width=1\linewidth]{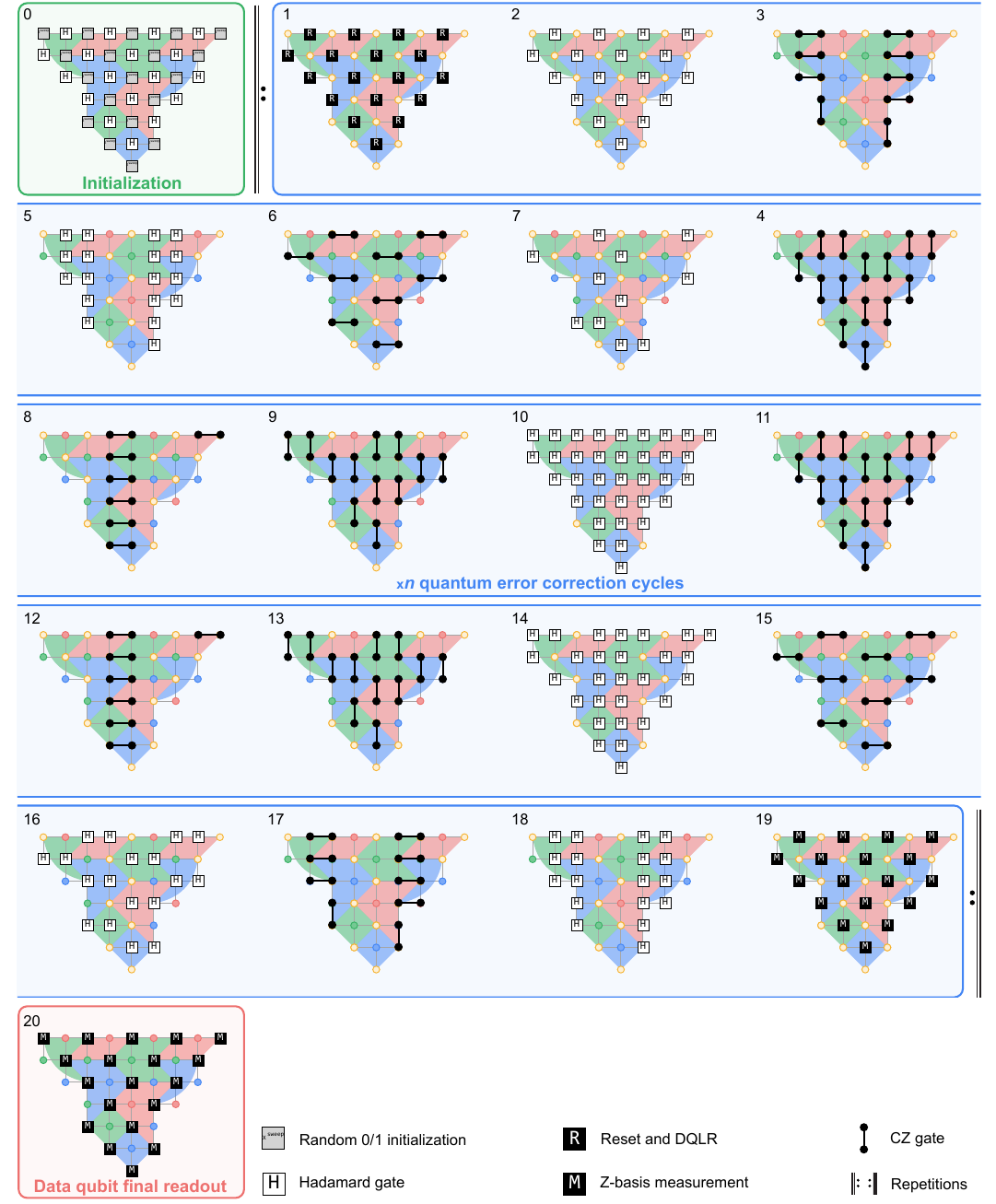}
    \caption{\textbf{Z-basis state preservation superdense color code circuit for the distance-five logical qubit.} The initialization, quantum error correction cycle and final data qubit readout are highlighted in green, blue, and red, respectively. See text for details.}
    \label{fig:superdense_cycle}
\end{figure*}

\section{Distance scaling} \label{app:distance_scaling}
In this Appendix, we provide additional details about the state preservation experiments performed on distance-three and distance-five logical qubits.
To mitigate effects from calibration drift, we interleave experiments with different codes, initial bases, and number of cycles following the approach described in Ref.~\onlinecite{Acharya2023}. 
Each experiment is repeated 5\,000 times for 10 different initialization bitstrings: five are randomly chosen and the remaining five are their binary complements, thereby ensuring equal representation of both eigenstates for odd-distance codes. This results in a total of 50\,000 experimental runs per experiment.

\subsection{Detailed Superdense Circuit}
We outline the steps of a Z-basis state preservation experiment on the distance-five logical qubit, after performing all circuit transformations described in \cref{app:circuit_transformation}. 
First, the data qubits are initialized in a randomly selected product state of Z-basis eigenstates (i.e.~a random bitstring) and all \auxiliary{} qubits are prepared in the $\ket{+}$ state (step 0 in \cref{fig:superdense_cycle}).
Then, we execute \nmemcycles{} cycles of error correction (steps 1 to 19).
During the \auxiliary{} qubit measurement and reset (steps 19 and 1), we apply an XY4 dynamical decoupling sequence~\cite{Maudsley1986} on all data qubits. After the \auxiliary{} qubit resets, we also apply data-qubit leakage removal (DQLR)~\cite{miao2023overcoming}.
Note that in the first cycle, resetting the \auxiliary{} qubits to the $\ket{+}$ state (frame 1 and 2) is skipped as it is already performed simultaneously to the data qubit initialization (step 0).
In the last cycle, the \auxiliary{} qubits are read out (step 19) simultaneously to all data qubits (step 20).

\subsection{Color Code Location on Processor} \label{app:color_code_location_on_processor}
\begin{figure*}
    \centering
    \includegraphics[width=\linewidth]{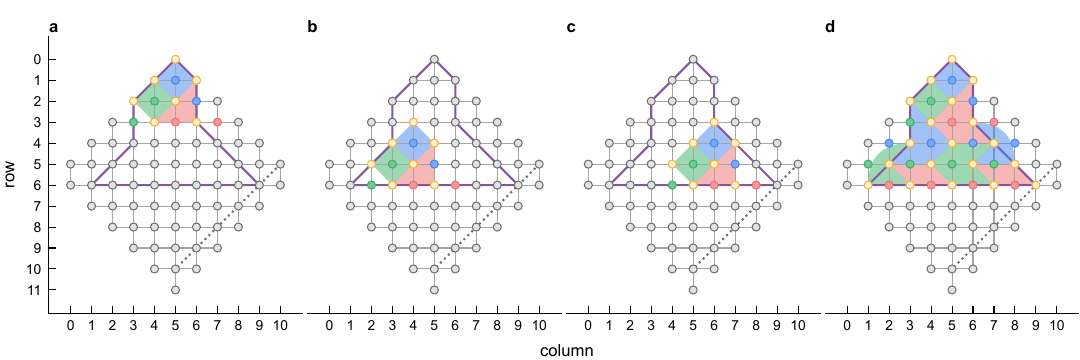}
    \caption{\textbf{Qubit subsets used to implement the three distance-three (\textbf{a}-\textbf{c}) and the single distance-five (\textbf{d}) color codes for the distance-scaling experiment}. Data qubits (\auxiliary{} qubits) for each subset are highlighted in gold (red, green, and blue), while inactive qubits of the 72-qubit processor are shown in gray. The purple outline marks the distance-five data qubit subset as a guide to the eye. Readout lines run diagonally from top right to bottom left, as indicated by a dotted gray line. }
    \label{fig:color_code_position_on_processor_grid}
\end{figure*}

We use 37 qubits of a 72-qubit quantum processor to encode the distance-five logical qubit, and three 13-qubits subsets to encode the distance-three logical qubits.  
The placement of the distance-three qubits on the processor is chosen to cover most of the qubits used in the encoding of the distance-five qubit, as visualized in \cref{fig:color_code_position_on_processor_grid}.

\subsection{Error syndrome detection probabilities} \label{app:detection_probabilities}
In each error correction cycle, the measurement of each \auxiliary{} qubit indicates the parity of its associated stabilizer: 0 for even parity and 1 for odd parity. 
After initialization, the stabilizer parity (and thus the \auxiliary{} qubit measurement outcome) should remain constant, except for intentional flips introduced by the circuit. 
Errors, such as bit-flips or phase-flips, alter the parity of neighboring stabilizers, leading to changes in the \auxiliary{} qubit measurement outcomes between consecutive cycles. Comparing  \auxiliary{} qubit measurements across cycles therefore reveals information about errors occurring on the device. 
These comparisons for individual \auxiliary{} qubits between two consecutive cycles are referred to as syndrome elements (also sometimes called detectors~\cite{Acharya2023}) and take a value of 1 if an error is detected by the syndrome extraction circuit, and 0 otherwise.

\begin{figure*}
    \centering
    \includegraphics[width=\linewidth]{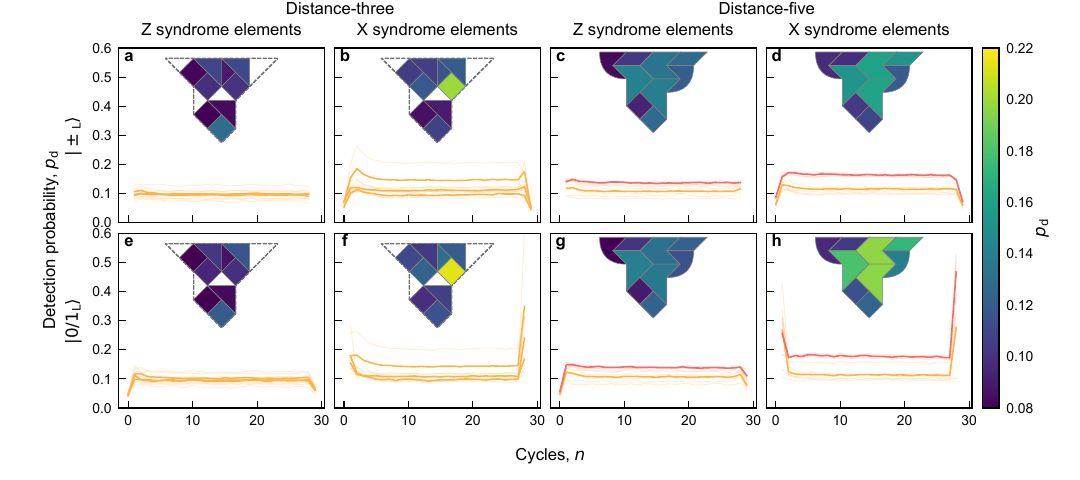}
    \caption{\textbf{Average error detection probabilities for all syndrome elements in the distance-scaling experiment}. 
    \textbf{a-d} Average weight-four (gold) and weight-6 (red) error detection probability, \pdet, as a function of the error correction cycle, \nmemcycles, in the $\ket{\pm_\mathrm{L}}$ (X-basis) state preservation experiments. 
    Panels \textbf{a} and \textbf{b} show the Z syndrome and the X syndrome detection probabilities of the distance-three logical qubits, respectively. Faded lines are the detection probabilities of individual syndrome elements while the solid lines correspond to their averages for each logical qubit. 
    Insets display heatmaps of the detection probabilities averaged over all cycles for each stabilizer of the code. The orientation of the code layouts in the insets matches the one of \cref{fig:color_code_concept} in the main text.
    Panel \textbf{c} and \textbf{d} display the Z syndrome and X syndrome detection probabilities for the distance-five logical qubit.    \textbf{e-h} is analogous to \textbf{a-d} but for the $\ket{0/1_\mathrm{L}}$ (Z-basis) state preservation experiments.   }
    \label{fig:det_fracs}
\end{figure*}

Analyzing how often each syndrome element detects an error provides insights into the frequency of errors across different parts of the device and stages of the quantum circuit. 
We calculate the error detection probability \pdet{} for each syndrome element by averaging its value across all experimental runs. For all codes and bases, syndrome elements in the bulk of the experiment ($5<\nmemcycles<25$) show relatively stable detection probabilities between 0.1 and 0.2, indicating that errors are rare and do not increase over time (see \cref{fig:det_fracs}).
In the distance-five state preservation experiments, weight-6 stabilizers (red lines) exhibit higher error detection probabilities than boundary weight-4 stabilizers (golden lines). This is  expected since weight-4 stabilizers involve fewer gates and are thus less prone to errors. 
Error detection probabilities are consistent across all experiments, except for certain X-type syndrome elements in the lowest distance-three code and the top-right part of the distance-five code, which have higher detection probabilities. Identifying the exact physical mechanism behind these additional phase errors requires further investigation. However, we hypothesize that they may result from spurious frequency shifts in data qubits induced by flux pulses used to reset \auxiliary{} qubits, which are not fully corrected by our in-situ phase calibration~\cite{Kelly2016}.
Near the time boundaries of the circuit ($\nmemcycles<5$ and $\nmemcycles>25$), higher error detection probabilities are observed in the X basis for both distance-three and distance-five codes.
We suspect this issue may be related to the one described above, and therefore warrants further investigation.

\section{Decoding} \label{app:decoding}
For each experimental run, the collection of all syndrome elements  form an error syndrome than can be used by a decoder to infer whether the logical operator has been flipped.
In the color code, each data qubit is included in three X-type and three Z-type stabilizers. Therefore, a single bit-flip or phase-flip error on a data-qubit in the bulk of the code will typically flip three syndrome elements (versus two for the surface code), creating a so-called hyper-edge in the syndrome graph. This prevents the direct use of the widely-used minimum-weight-perfect-matching decoders~\cite{higgott2023sparse}, which rely on each error triggering a pair of syndrome elements. 

In this work, we explore three decoders for the color code: a möbius decoder (Chromobius)~\cite{Gidney2023}, a most-likely-error (MLE) decoder, and a neural-network decoder (AlphaQubit)~\cite{Bausch2024}.
We provide a short description of each of these decoders in the following Subsection and highlight their advantages and limitations.

\subsection{Decoder details}
\subsubsection{Möbius Decoder (Chromobius)}
Chromobius~\cite{Gidney2023}  decomposes the color-code decoding problem in three main steps.
First, it decomposes the color-code decoding graph into three complementary matchable graphs without hyper-edges. These subgraphs are connected at their spacial boundaries in a Möbius topology~\cite{Sahay2022}. Compared to other decoders that use similar decomposition methods~\cite{Kubica2023}, connecting the boundaries enables more accurate weighting of the relative costs of matching syndrome elements to a boundary (typically underestimated when graphs are disconnected) versus matching them to other syndrome elements in the bulk of the complementary subgraphs~\cite{Sahay2022, Gidney2023}.
The resulting decoding graph is then processed by a minimum-weight perfect matching algorithm to find a perfect matching solution.  
Finally, the solution is lifted back to the original color code decoding graph by merging the matched syndrome elements using a dynamic programming algorithm~\cite{Gidney2023}. 

Chromobius decodes data much faster than many other color code decoders, including the two other decoders used in this work, because the graph decomposition, the matching subroutine, and the solution lifting steps can all be performed efficiently. This makes Chromobius a promising candidate for real-time decoding in the near future.

Nonetheless, Chromobius has two notable limitations. First, it does not currently support flag-based decoding nor the ability to exploit correlations between the Z-basis and X-basis matching graphs. 
Second, it does not support certain logical operations that modify the decoding graph, such as those transforming X stabilizers into Y stabilizers. 
Consequently, Chromobius does not provide optimal performance and cannot be used to decode logical randomized benchmarking circuits.

In this work, we use Chromobius for the state injection experiment, and compare its performance to other decoders for the state preservation experiments (see \cref{app:decoding_comparison} for details).

\subsubsection{Most Likely Error (MLE) Decoder}
Inspired by Refs.~\onlinecite{Landahl2011, Takada2024}, we develop a most-likely-error (MLE) decoder that identifies the most likely chain of Pauli errors given the observed error syndrome. The most-likely error is found by mapping the decoding problem onto a mixed-integer program~\cite[Supp. 1A]{Aggoun2008, Cain2024} which is solved with the High Performance Optimization Software (HiGHS) package~\cite{Huangfu2018}. This method is more accurate than matching-based approaches and naturally accounts for correlations between X and Z syndrome elements. 
Additionally, the Pauli error chains returned by the decoder can be used to iteratively refine the decoder's prior knowledge of frequent errors, as detailed in \cref{app:decoding_priors}. 

The MLE decoder also has limitations. 
First, it is considerably slower than the other decoders (currently about 1000 times slower than the Möbius decoder and 300 times slower than the neural-network decoder used in this work).
Second, while being more accurate than matching-based decoders, it is not an optimal decoder such as a maximum likelihood decoder~\cite{Sundaresan2023} or its tensor-network approximation~\cite{Acharya2023}. These optimal decoders consider the probability of all possible error sets yielding the observed error syndrome, dividing them into two groups: those that flip the logical operator and those that do not. They then select the group with the greater total likelihood. However, such optimal decoders require much larger computational resources than our MLE decoder.

We compare the performance of the MLE decoder to other decoders for the state preservation experiments and use it in the simulations described in \cref{app:distance_scaling_simulations,app:threshold_simulations}.

\subsubsection{Neural Network Decoder}
AlphaQubit is a recurrent attention-based neural network first described in Ref.~\onlinecite{Bausch2024}, and used in Ref.~\onlinecite{Acharya2024} to decode surface code experiments.
Here, we adapt the network to decode color code experiments. Because of the lower translational symmetries of the color code compared to the surface code, the architecture of the model does not contain convolutions. In all experiments decoded in this work besides the state-teleportation experiments, we add up to two attention/recurrent layers to compensate for the loss of computational power. The resulting networks are significantly smaller: 4.1 million parameters (for the two layer variant) and 2.1 million (for the single layer variant), compared to the 5.4 million used in Ref.~\onlinecite{Acharya2024}. We use a recurrence update as described in Ref.~\onlinecite{Bausch2024}, rather than a gated-recurrence~\cite{Cho2014} update as in Ref.~\onlinecite{Acharya2024}.

For each experiment, AlphaQubit is trained to predict the logical observable, based on the error syndrome calculated from the stabilizer measurements. The network processes one cycle of syndrome elements at a time, and (for the randomized benchmarking experiments) receives information about logical gates applied in between cycles, updating an internal state representation which consists of a vector for each stabilizer. At the end of the experiment the state vectors are combined and further processed to make the prediction of the logical observable. 

For state preservation experiments, separate models are trained for the X and Z bases for each code instance. By contrast, for logical randomized benchmarking and state-teleportation experiments, a single model is used for all logical circuits in each respective experiment.
For state teleportation, a single instance of AlphaQubit processes the syndrome elements from both logical qubits in each cycle, along with additional syndrome elements corresponding to the stabilizers at the interface of the logical qubits during the merge cycles.  

The network is trained in two stages. First, it is pre-trained on synthetic data generated with Stim~\cite{Gidney2021} using a device-agnostic noise model (SI1000 with a range of noise levels). 
After pretraining, each network is fine-tuned by further training on a limited amount of experimental data (or device-specific simulated data e.g.~for the error budget). We do not fine-tune the network on a synthetic dataset generated with an error model derived from experimental data (as was previously done in Ref.~\onlinecite{Acharya2024}). 
Particular care is taken to never train and test on the same data, by either using: 
\begin{itemize}
\item[--]{an even/odd experimental run partitioning (randomized benchmarking),}
\item[--]{a causal split where we only fine-tune on data taken in a separate run prior to the test dataset (state preservation experiments),}
\item[--]{a fully independent synthetic training dataset (error budget), or}
\item[--]{a separate part of the experimental dataset for testing purposes (state teleportation).}
\end{itemize}

After training, AlphaQubit can decode at an intermediate speed which is about 3 times slower than Chromobius, but 300 times faster than the MLE decoder. Due to its high accuracy, AlphaQubit is the decoder we used used in most of the main text of this article (besides the magic state injection experiments, where we make use of Chromobius and post-selection).

\subsection{Decoding Priors}\label{app:decoding_priors}
Both Chromobius and the MLE decoder rely on prior knowledge of the probabilities of individual errors occurring in the circuit. To provide this information, we explore three different approaches.

The simplest method calculates error probabilities of each operation in the circuit based on the SI1000 error model~\cite{Gidney2021a} parametrized with a single error parameter $p$. This model reflects the typical hierarchy of error rates in superconducting qubits (for instance, all two qubit gates have an error $p$ while single-qubit gates have an error $p/10$) but remains device-agnostic.

The remaining two approaches make use of experimental data to refine a Pauli error model where each circuit operation has an associated a depolarizing error parameter and/or a bit-flip error parameter. The first experimentally-informed approach derives error probabilities for each circuit operation from experimental data decoded with Chromobius, following the reinforcement-learning methodology outlined in Ref.~\onlinecite{sivak2024optimization}. The error probabilities are
inferred from the 9-cycles dataset for each experiment separately, and then extrapolated using time-invariance to all other numbers of cycles~\cite{sivak2024optimization}.

Finally, we explore an edge-reweighting strategy inspired by Ref.~\onlinecite{Wang2023}, which iteratively adjusts the probabilities of individual errors based on their frequency as reported by the MLE decoder. Initially, all errors are assumed to be equally likely. 
In the first iteration, a calibration dataset is decoded using the MLE decoder, which provides the most likely error chain for each experimental run in the dataset. 
The probabilities of individual errors are then updated by counting their occurrences in the list of most likely error chains across the dataset. 
This process is repeated until the decoded performance stabilizes, typically after 3 to 6 iterations. 

\subsection{Comparison of Decoding  Performance}\label{app:decoding_comparison}

\begin{figure}
    \centering
    \includegraphics{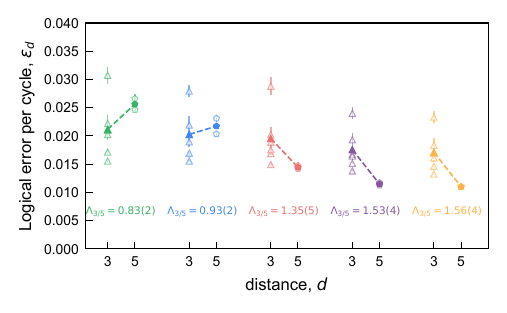}
    \caption{\textbf{Comparison of decoding strategies for the distance scaling experiment.} Logical error per cycle of the state preservation experiments as a function of code distance for five decoding strategies: Chromobius decoder with SI1000 prior (green), Chromobius decoder with reinforcement-learning prior (blue), MLE decoder with SI1000 prior (red), MLE decoder with iterative edge-reweighting prior (purple), neural network decoder fine-tuned on experimental data (yellow). Symbols with contours only correspond to the logical error rates for individual experiments (i.e.~one code instance in a given basis) and the filled symbols correspond to the average over all experiments. Error bars indicate the standard error from the fit. The dashed line serves as a guide to the eye. The error suppression factor \errorsuppressionfactor{} is indicated for each decoding strategy with the corresponding color.}
    \label{fig:decoder_comparison}
\end{figure}

We use the state preservation experiments on the distance-three and distance-five codes to compare five decoding strategies: 
\begin{enumerate}
    \item Chromobius decoder with an SI1000 prior (green symbols in \cref{fig:decoder_comparison})
    \item Chromobius decoder with a learned prior from experimental data (blue symbols)
    \item MLE decoder with an SI1000 prior (red symbols)
    \item MLE decoder with a prior obtained via iterative edge reweighting based on experimental data (purple symbols)
    \item Neural-network decoder (AlphaQubit), pretrained on SI1000 simulated data and fine-tuned on experimental data (yellow symbols)
\end{enumerate}

For each decoding strategy, we fit the decoded data and extract the logical error per cycle for each code instance and basis independently. Note that, for clarity, we show the logical error as a function of cycle number and the logical error per cycle averaged over the X and Z bases for each code instance in \cref{fig:distance_scaling} of the main text. 
We average all distance-three and all distance-five experiments, and compute the error suppression factor \errorsuppressionfactor. We use the standard error reported by each fit and standard error propagation techniques to obtain error bars for the error suppression factor. 

Three of the five strategies achieve a lower average logical error per cycle for the distance-five code compared to the distance-three code, as shown in \cref{fig:decoder_comparison}. 
The best decoding strategy provides an error suppression factor of $\errorsuppressionfactor = \lambdabest$, demonstrating operation of the color code below its critical noise threshold. 
As expected, both the Chromobius and MLE decoders perform better with experimentally-informed priors than with device-agnostic (SI1000) priors. 
Notably, the MLE decoder achieves performance comparable to the neural-network decoder, highlighting that a simple Pauli error model with depolarizing parameters for each circuit operations captures well the errors of the device. 
However, the neural-network decoder is significantly faster than the MLE decoder.

Note that the three decoding strategies relying on experimental data (strategies 2, 4, and 5) use different train/test data partitioning schemes. Consequently, the test dataset for each of these strategies may contain different experimental runs.

\section{Pauli Simulations near the Superdense Color Code Threshold}\label{app:threshold_simulations}
In this article, we observe clear error suppression ($\errorsuppressionfactor = \lambdabest$) when scaling the distance of the color code from three to five.
However, there is a crossover region near the code threshold ($\Lambda \approx 1$) where the error suppression factor $\Lambda$ can depend on the code distance, as reported in Ref.~\onlinecite{Acharya2023}.
Therefore, to relate our experimental results to the color code threshold, we perform simulations extrapolating to larger code distances.

\begin{figure*}
\includegraphics{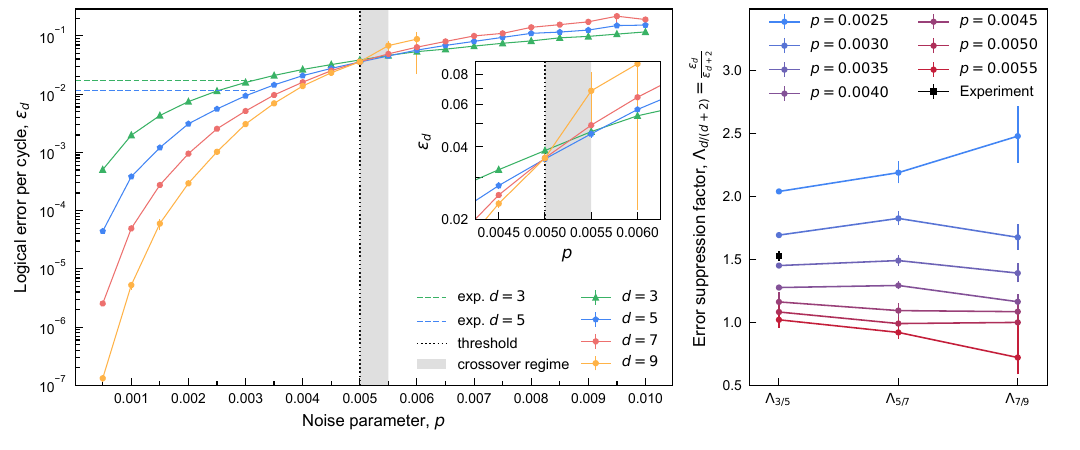}
    \caption{\textbf{Simulation of the superdense color code using an SI1000 error model and decoded with the MLE decoders.} \textbf{a}, Logical error per cycle versus the noise parameter $p$ of the SI1000 error model. The different curve correspond to different code distances $d$, and the gray region represents an approximate crossover region.  We define the threshold (dotted black line) as the left bound of this crossover region.  For smaller $p$ than the $\pth\approx 0.005$, we find that higher distances outperform lower distances with a confidence interval of 90\,\% (which is represented by the error bars). 
    The horizontal dashed lines correspond to the experimental value for \meanepc[3] (green) and \epc[5] (blue). \textbf{b,} Simulated $\Lambda_{d, d+2}$ for different values of $d$. Each curve is associated with a different $p$ noise strength, and the black point corresponds to the experimentally measured \errorsuppressionfactor{}.
    }
    \label{fig:below_threshold_simulations}
\end{figure*}

We simulate the superdense color code circuit from Ref.~\onlinecite{Gidney2023} for distances $d=3$ to $9$ using an SI1000 error model~\cite{Gidney2021a} with varying noise strength $p$, and decode the data with the MLE decoder.
We observe that the threshold at which the logical error per cycle for the higher distances intersect is around $\pth\approx0.005$ (see \cref{fig:below_threshold_simulations}).
Near this threshold, \errorsuppressionfactor{} is about 1.1 but decreases to about 1.0 at higher distances $\Lambda_{d/(d+2)}$.
The simulated logical error per cycle for $p=0.005$ is $\epc[3] = 0.039(1)$ and $\epc[5] = 0.036(1)$.

We compare these simulated values to the experimental observations.
Experimentally, we observe $\errorsuppressionfactor{}=\lambdasimplex$, $\meanepc[3]=\meanepcthreesimplexvalue$ and $\meanepc[5] = \epcfivesimplexvalue$, using the same MLE decoder. These are significantly lower than the simulated values at threshold, indicating that our experiment is indeed below the color code noise threshold.

Note that the SI1000 error model is a simplified representation of the experimental system; it assumes homogeneity and does not account for error mechanisms such as leakage and stray interactions.
Therefore, it would be valuable to experimentally measure the logical error per cycle at larger code distances in the future.

\section{Logical Randomized Benchmarking} \label{app:logical_rb}
Demonstrating the implementation of logical Clifford gates is a critical step for fault-tolerant quantum computing, but benchmarking their error rates at the logical level presents significant challenges. 
Previous experimental efforts have primarily relied on partial~\cite{Postler2022} or full~\cite{Marques2022} process tomography. However, this method is sensitive to logical SPAM (State Preparation and Measurement) errors, which can be comparable to or even exceed the gate error itself, especially for the surface code in which Y-basis readout is not fault-tolerant~\cite{Marques2022}. 
Additionally, it may not accurately capture the performance of gates when applied within a logical circuit. Similar to how randomized benchmarking~\cite{Magesan2012} provides robust error estimates at the physical qubit level, logical randomized benchmarking~\cite{Combes2017} offers a more reliable method for assessing the cost of logical Clifford gates within logical circuits.

We focus on the implementation of LRB on a single distance-three qubit. The 24 single-qubit logical Clifford gates can be decomposed into a sequence of Hadamard gates $H$ ($\pi$-rotation along the $x+z$ axis of the Bloch sphere), phase gates $S$ ($\pi/2$-rotation along the $z$ axis of the Bloch sphere) and Pauli $X, Y,$ and $Z$ gates.
Specifically, they are obtained from a matrix product $A\cdot B$ where $A \in \{I, H, S, HS, SH, HSH\}$ and $B = \{I, X, Y, Z\}$.
In our implementation, we first generate logical circuits comprising a sequence of \nrclifford{} randomly chosen Clifford gates, followed by a Clifford recovery gate~\cite{Magesan2012} that returns the logical qubit to its initial state (\zerological{} in our case). These gates are interleaved with error correction cycles. 
The circuits are then compiled into individual gates compatible with our hardware. During this compilation, single-qubit gates used for transversal logical gates may be merged with single-qubit operations from the neighboring error correction cycles. This approach is reasonable because it could be employed for more complex logical circuits.
The observed logical gate errors reflect not only by the additional physical operations performed for logical gates but also by the frequent basis changes on the logical qubit, including time spent in the Y-basis, which is vulnerable to both bit-flip and phase-flip errors.

Using the idling experiment as reference~\cite{Magesan2012}, we extract the additional error introduced by logical Clifford gates per cycle
\begin{equation}
    \errorperclifford = \frac{1}{2} (1 - \frac{\depolarizingpclifford}{\depolarizingpidle})
\end{equation}
where \depolarizingpclifford{} and \depolarizingpidle{} are the depolarizing parameters of the decaying Clifford experiment and of the reference idling experiment, respectively. 
\depolarizingpclifford{} and \depolarizingpidle{} are obtained from fitting the average values shown in \cref{fig:logical_rb}e of the main text to the function $A p^{\nrclifford{}} + 1/2$ with fitting parameters $A$ and $p\in \{\depolarizingpclifford, \depolarizingpidle\}$.

Note that LRB is a benchmarking tool, and the logical circuits it generates are not intended for practical applications, as these circuits comprising of Clifford gates only can be efficiently simulated on a classical computer~\cite{Gottesman1998}.

\section{State injection} \label{app:state_injection}

\begin{figure}
    \centering
    \includegraphics[width=1\linewidth]{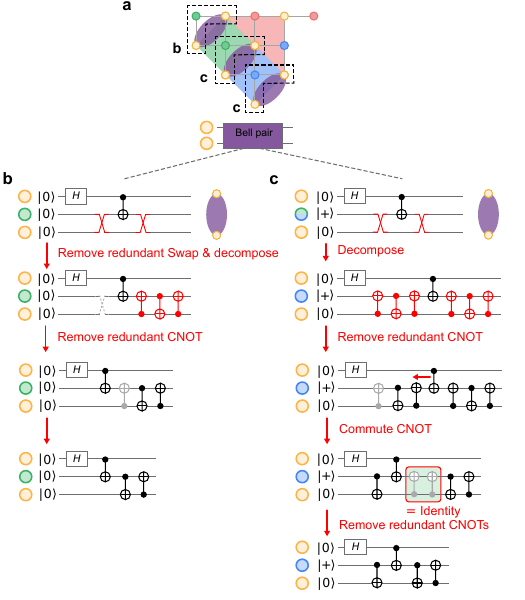}
    \caption{\textbf{Bell state initialization circuit for magic state injection}.
    \textbf{a} Schematic of a distance-three color code, with data qubits shown as gold dots and \auxiliary{} qubits as red, blue, and green dots. Gray lines represent qubit connectivity. Purple ellipses indicate pairs of data qubits initialized in Bell states for the magic state injection protocol (see main text for details). The dashed black contour highlights qubit subsets used to generate Bell states with the circuits shown in \textbf{b} and \textbf{c}.
    \textbf{b} Circuit transformations to generate a Bell state between two data qubits using an intermediate \auxiliary{} qubit initialized in the $\ket{0}$ state. Red arrows indicate the sequence of circuit transformations. Starting with a circuit using swap gates (highlighted in red), the first swap gate is removed as both qubits are in the $\ket{0}$ state. The remaining swap gate is decomposed into three CNOT gates (second circuit), with the first CNOT removed (grayed out in third circuit) since its control qubit is in $\ket{0}$. The resulting circuit uses 3 two-qubit gates (fourth circuit). \textbf{c} Similar approach as in \textbf{b} but for a set of qubits in which the \auxiliary{} qubit initialized in $\ket{+}$. The first swap gate cannot be removed trivially. Swap gates are decomposed into CNOT gates (second circuit), the first CNOT is removed since its control qubit is in $\ket{0}$, and commuting CNOT gates are reordered (third circuit). Two consecutive CNOT gates cancel out (fourth circuit), resulting in a final circuit with 4 two-qubit gates (fifth circuit). }
    \label{fig:state_injection_bell_pair_circuit}
\end{figure}

To inject an arbitrary state \arbitrarymagicstate{} into a distance-three logical qubit, we first prepare the desired state on a physical qubit and then use a fault-tolerant protocol~\cite{Zhang2024, Jones2016} to expand the color code from a degenerate distance $d=1$ code to a $d=3$ code. 
In this protocol, the six data qubits that are added to the code are pairwise initialized in Bell states, as indicated by the purple ellipses in \cref{fig:state_injection_bell_pair_circuit}a.
The standard circuit to prepare a Bell state from two neighboring qubits initially in the zero state involves applying a Hadamard gate to the first qubit, followed by a CNOT gate with the first qubit acting as the control. 
However, since these pairs of data qubits are not directly connected on our device, we create the Bell state between them using an intermediate \auxiliary{} qubit, as indicated by the dashed black contour in \cref{fig:state_injection_bell_pair_circuit}a. 
This process requires three two-qubit gates if the \auxiliary{} qubit starts in the $\ket{0}$ state, or four if it begins in the $\ket{+}$ state, see \cref{fig:state_injection_bell_pair_circuit}b and \cref{fig:state_injection_bell_pair_circuit}c for a visual derivation of the circuit, respectively.
While the Bell pairs are prepared, the injection qubit is prepared in a desired arbitrary state with single-qubit gates parametrized by the polar and azimuthal angles introduced in the main text.

After this initialization step, we perform a standard error correction cycle on the distance-three qubit, followed by a logical measurement in the X, Y, or Z basis for logical state tomography.
We repeat the circuit 20\,000 times for each basis and compute the expectation value of the corresponding logical operator $\langle O_\mathrm{L}\rangle = 1 - 2 \langle m_\mathrm{O}\rangle $, where $\langle m_\mathrm{O}\rangle$ is the logical outcome averaged over all repetitions. 

\begin{figure}
    \centering
    \includegraphics[]{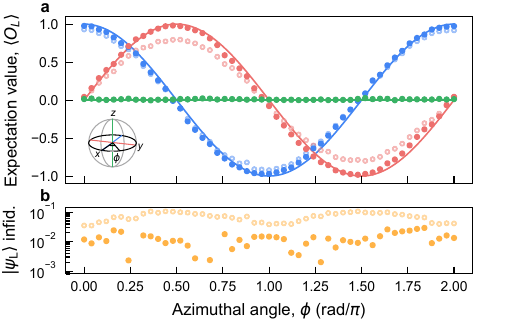}
    \caption{\textbf{Arbitrary state injection along the $x$-$y$ plane of the logical Bloch sphere.} \textbf{a} Decoded (semi-transparent circles) and post-selected (solid dots) expectation value of the logical Pauli operators \lopx{} (blue), \lopy{} (red), \lopz{} (green) when sweeping the azimuthal angle \azimuthalangle. The solid lines correspond to ideal expectation values. \textbf{d} Decoded (semi-transparent circles) and post-selected (solid dots) infidelities for the prepared logical state \arbitrarymagicstate.}
    \label{fig:state_injection_xy_sweep}
\end{figure}

We show the expectation values of \lopx, \lopy, and \lopz{} for states along $x$-$y$ plane of the Bloch sphere in \cref{fig:state_injection_xy_sweep}a. To prepare these states on the injection qubit, we fix the polar angle $\polarangle=\pi/2$ and sweep the azimuthal angle \azimuthalangle{} from 0 to $2\pi$. Note that we add a phase offset of $0.044\pi\approx 7.9^\circ$ to compensate for a coherent phase rotation of the logical qubit which we measured in a previous calibration measurement.
We reconstruct the approximate density matrix 
\begin{equation}
    \tilde{\rho}_\mathrm{L} = \frac{1}{2} \left(I + \langle X_\mathrm{L}\rangle \sigma_x + \langle Y_\mathrm{L}\rangle \sigma_y + \langle Z_\mathrm{L}\rangle \sigma_z  \right)
\end{equation}
where $I$ is the identity matrix and $\sigma_{x,y,z}$ are the Pauli matrices. 
We then calculate the state fidelity 
\begin{equation}
    \mathcal{F}(\tilde{\rho}_\mathrm{L}, \rho_\mathrm{L}) = \left(\trace\sqrt{\sqrt{\tilde{\rho}_\mathrm{L}} \hspace{0.5em} \rho_\mathrm{L} \sqrt{\tilde{\rho}_\mathrm{L}} }\right)^2
\end{equation} 
to the target density matrix $\rho_\mathrm{L} = \arbitrarymagicstate\bra{\psi_\mathrm{L}}$, where $\trace$ denotes the trace operation.

We obtain an average infidelity  of  $1-\mathcal{F}(\tilde{\rho}_\mathrm{L}, \rho_\mathrm{L})  =\azimuthalangleinfidpostselectionvalue{}$ over the sweep when post-selecting on experimental runs in which all syndrome elements are 0 (keeping \azimuthalanglepostselectionfraction{} of the data), see \cref{fig:state_injection_xy_sweep}b. 
We observe the largest discrepancy between decoded and post-selected values for \lopy, which we attribute to its sensitivity to both bit-flip and phase-flip errors.

As described in the main text, we also prepare specific magic states of interest for realizing $T$-gates. For these states, we perform an additional numerical optimization to obtain the maximum likelihood estimate of the experimentally reconstructed density matrix which ensures physicality constraints (all positive eigenvalues and a trace equal to 1).

\begin{figure}
    \centering
    \includegraphics[]{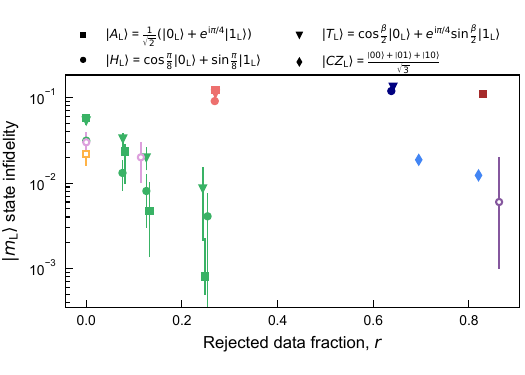}
    \caption{\textbf{Comparison of magic state injection experiments.} Infidelity of four experimentally realized magic states (see legend) plotted against the rejected data fraction, $r$, with $r=0$ corresponding to no post-selection. Open symbols represent implementations with trapped ions, while filled symbols correspond to superconducting qubits. Each color represents states realized in a different study (see text for references). Some works (including ours) report results with various post-selection strategies, leading to several symbols for the same magic state to appear multiple times in the same color. All data points have error bars, but some are too small to see. The error bars reported from our work correspond to a 95\,\% confidence interval calculated using bootstrapping.}
    \label{fig:state_injection_comparison}
\end{figure}

We compare our results to other experimentally prepared magic states in prior work with superconducting qubits~\cite{Marques2022, Ye2023, Gupta2024, Kim2024} and trapped ions~\cite{Egan2021, Ryan-Anderson2021, Postler2022}, see \cref{fig:state_injection_comparison} for an overview of magic state infidelity as a function of the rejected data fraction. 
We observe that our implementation (green dots) can achieve higher fidelities and lower data rejection fractions than other implementations with superconducting qubits, and is on the Pareto front with some trapped ion implementations.

\section{State teleportation} \label{app:state_teleportation}
\subsection{State Teleportation Circuit Details}
\begin{figure*}
    \centering
    \includegraphics[]{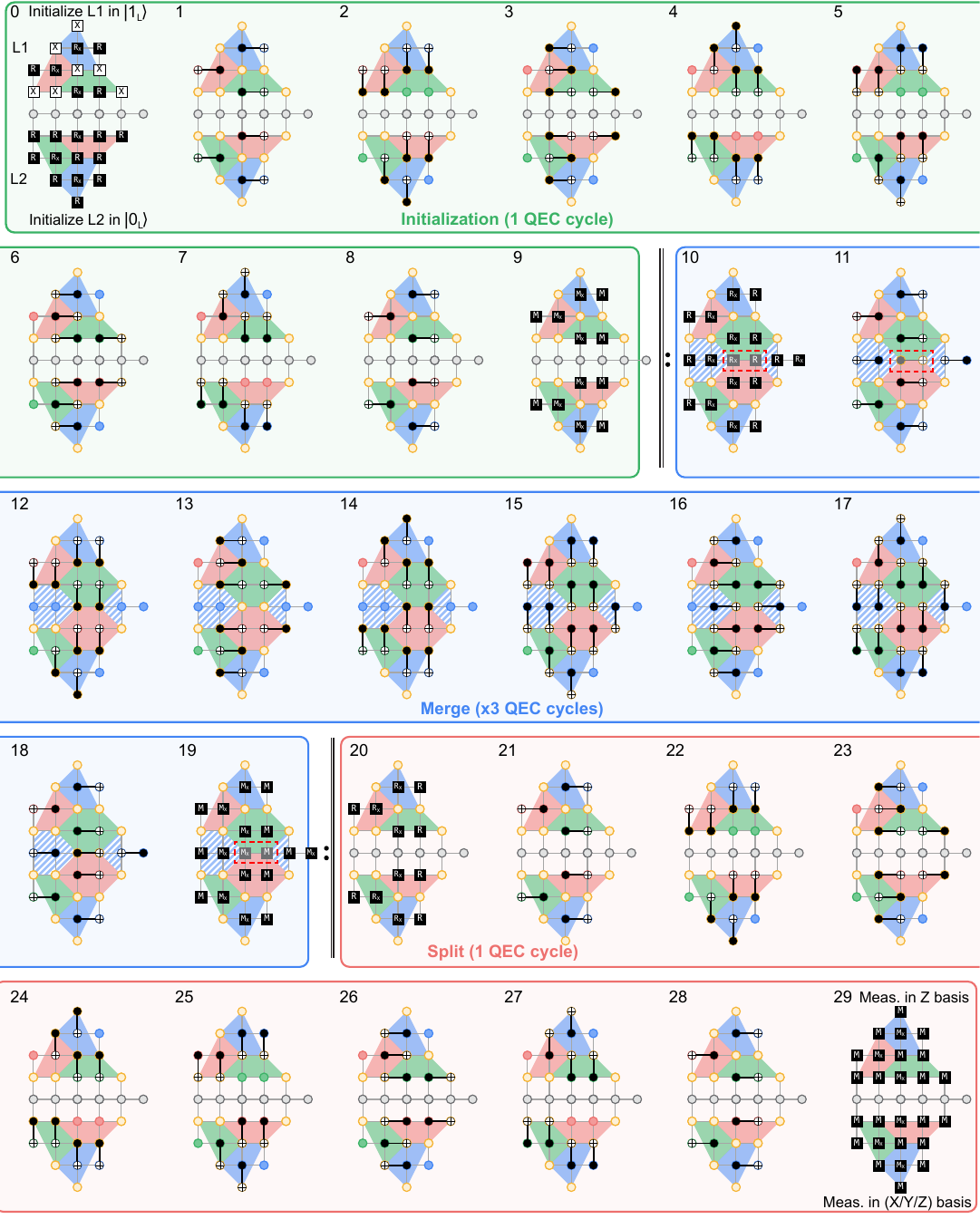}
    \caption{\textbf{Detailed quantum circuit used to teleport \onelogical{} from logical qubit L1 to logical qubit L2.} Active data qubits (\auxiliary{} qubits) are denoted with golden dots (red, green and blue dots), and inactive qubits are shown in gray. The red, green, and blue tiles indicate X and Z stabilizers with support on the data qubits located at the vertices of the tile, while the hatched blue tiles indicate X stabilizers only. The green, blue, and red boxes highlight the three parts of the protocol: initialization, merge, and split. Double black lines indicate repeated cycles of the merge operation, with dashed red outlines marking operations performed only in the first or last cycle (see text for details). The black boxes labeled R and M (Rx and Mx) correspond to reset and measurement operations in the Z basis (X basis), respectively. }
    \label{fig:state_teleportation_detailed_circuit}
\end{figure*}
We provide a detailed circuit for teleporting \onelogical{} from logical qubit L1 to logical qubit L2 using lattice surgery. The circuit is divided into three parts: initialization, \mxx{} merge (X basis parity measurement of the two logical qubits), and split, highlighted in green, blue, and red in \cref{fig:state_teleportation_detailed_circuit}, respectively.

During initialization, one cycle of error correction is performed separately on each logical qubit. The merge operation introduces stabilizers in the X basis at the interface between the two logical qubits (hashed blue tiles) and expands two stabilizers from weight-4 to weight-6 by incorporating two additional data qubits. In the first merge cycle, these interface data qubits are prepared in a Bell state (grayed operation in the dashed red box, panels 10-11) and are measured in the Bell basis during the third merge cycle (grayed operation in the dashed red box, panel 19). The \mxx{} parity outcome is the product of the two interface X stabilizers and the additional data qubit read out in the X basis.

The split operation consists of a single error correction cycle, after which the data qubits of L1 are measured in the Z basis, while those of L2 are measured in the X, Y, or Z basis for logical tomography. Pauli frame updates are applied to the logical operators of L2 based on the merge outcome and the measurement result of qubit L1, as outlined in the circuit diagram in \cref{fig:state_teleportation}a in the main text.

\subsection{Lower Bound on Average Fidelity of the Teleportation Channel}
In the main text, we report the expectation values of logical Pauli operators and the corresponding state fidelity for the teleportation of the Pauli eigenstates \zerological, \onelogical, \pluslogical, and \minuslogical. 
These values can be used to provide a lower bound on the \textit{average fidelity} \teleportationaveragefidelity{}~\cite{Nielsen2002} of the teleportation channel \teleportationchannel{} with respect to the identity channel \identitychannel. 

First, we note that~\cite{Nielsen2002} 
\begin{equation} \label{eq:average_fidelity}
    \teleportationaveragefidelity(\teleportationchannel) = \frac{d\teleportationentanglementfidelity(\teleportationchannel) + 1}{d+1}
\end{equation}
where $d=2$ is the dimension of the single logical qubit Hilbert space and 
\begin{equation} \label{eq:entanglement_fidelity}
    \teleportationentanglementfidelity(\teleportationchannel) = \langle\phi|\left(\mathcal{I}\otimes\teleportationchannel\right)(|\phi\rangle\langle\phi|)|\phi\rangle
\end{equation}
is the \textit{entanglement fidelity}~\cite{Schumacher1996} of the teleportation channel. 
Here, $|\phi\rangle=(|0_\mathrm{L}0_\mathrm{L}\rangle+|1_\mathrm{L}1_\mathrm{L}\rangle)/\sqrt{2}$ is a Bell state of two logical qubits. 
Suppose $\teleportationchannel$ has a Kraus representation $\teleportationchannel(\rho)=\sum_k K_k \rho K_k^\dagger$. By expanding each Kraus operator in the logical Pauli basis, we can express $\teleportationchannel$ as
\begin{align}
    \teleportationchannel(\rho) &= \sum_k K_k \rho K_k^\dagger
    \\
    &= \sum_k\left(\sum_i a_{ki}P_i\right) \rho\left(\sum_j a_{kj}P_j\right)^\dagger \\
    &= \sum_{ij}\left(\sum_ka_{ki} \overline{a_{kj}}\right) P_i \rho P_j
\end{align}
where $i,j\in\{0,1,2,3\}$, $P_0=I_\mathrm{L}$, $P_1=X_\mathrm{L}$, $P_2=Y_\mathrm{L}$, $P_3=Z_\mathrm{L}$, and $k\in\{1,\dots,r\}$ where $r$ is the channel's Kraus rank. The matrix $\chi_{ij} = \sum_k a_{ki} \overline{a_{kj}}$ is called the channel's \textit{process matrix}~\cite{Nielsen2010}. The action of $\teleportationchannel$ on a density matrix $\rho$ can then be written in terms of the process matrix as
\begin{align} \label{eq:process_matrix}
    \teleportationchannel(\rho)=\sum_{ij}\chi_{ij}P_i\rho P_j.
\end{align}
Note that $\chi$ is positive semidefinite and has unit trace. Moreover,
\begin{align}
    F(\teleportationchannel) &= \sum_{ij}\chi_{ij}\langle\phi|\left(I\otimes P_i\right)|\phi\rangle\langle\phi|\left(I\otimes P_j\right)|\phi\rangle \\
    &= \frac{1}{4}\sum_{ij}\chi_{ij}\mathrm{tr}(P_i)\mathrm{tr}(P_j) \\
    &= \chi_{00}
\end{align}
where the first equality results from the substitution of \cref{eq:process_matrix} into \cref{eq:entanglement_fidelity} and the second from the definition of $|\phi\rangle$. Therefore, we can read entanglement fidelity directly from a single element of the process matrix $\chi$.

For any two (logical) Pauli operators $P$ and $Q$, the quantity $\frac{1}{2}\mathrm{tr}(Q\teleportationchannel(P))$, which we term \textit{Pauli correlation}, is experimentally accessible by taking linear combinations of preparations of the eigenstates of $P$ and measuring in the basis of $Q$. To see this, suppose that we perform the following experiment $N^0$ times: prepare $|0_\mathrm{L}\rangle$, apply $\teleportationchannel$ and measure the observable $Z_\mathrm{L}$, obtaining the $+1$ eigenvalue $N^0_\text{agree}$ times and the $-1$ eigenvalue $N^0_\text{disagree}$ times with $N^0_\text{agree} + N^0_\text{disagree} = N^0$. Then the Born rule implies that
\begin{align} \label{eq:experiment_with_0L_preparation}
    \mathrm{tr}(Z_\mathrm{L}\teleportationchannel(|0_\mathrm{L}\rangle\langle 0_\mathrm{L}|)) = \frac{\mathbb{E}[N^0_\text{agree}] - \mathbb{E}[N^0_\text{disagree}]}{N^0}
\end{align}
where $\mathbb{E}$ denotes the expectation value. Similarly, if we perform a similar experiment $N^1$ times, preparing $|1_\mathrm{L}\rangle$ instead of $|0_\mathrm{L}\rangle$ and obtain the $+1$ eigenvalue $N^1_\text{disagree}$ times and the $-1$ eigenvalue $N^1_\text{agree}$ times with $N^1_\text{agree} + N^1_\text{disagree} = N^1$, then
\begin{align} \label{eq:experiment_with_1L_preparation}
    \mathrm{tr}(Z_\mathrm{L}\teleportationchannel(|1_\mathrm{L}\rangle\langle 1_\mathrm{L}|)) = \frac{\mathbb{E}[N^1_\text{disagree}] - \mathbb{E}[N^1_\text{agree}]}{N^1}.
\end{align}
Subtracting \cref{eq:experiment_with_1L_preparation} from \cref{eq:experiment_with_0L_preparation} and using linearity of the trace and the channel $\teleportationchannel$, we find
\begin{align}
    \mathrm{tr}(Z_\mathrm{L}\teleportationchannel(Z_\mathrm{L}))
    =& \frac{\mathbb{E}[N^0_\text{agree}] - \mathbb{E}[N^0_\text{disagree}]}{N^0} +\\
    & \frac{\mathbb{E}[N^1_\text{agree}] - \mathbb{E}[N^1_\text{disagree}]}{N^1}.
\end{align}

We can also express Pauli correlations as linear combinations of the elements of the process matrix. In particular,
\begin{align}
    \frac{1}{2}\mathrm{tr}(X_\mathrm{L}\teleportationchannel(X_\mathrm{L})) 
    &= \frac12\sum_{ij}\chi_{ij}\mathrm{tr}(XP_iXP_j) \\
    &=\chi_{00}+\chi_{11}-\chi_{22}-\chi_{33}
\end{align}
and
\begin{align}
    \frac{1}{2}\mathrm{tr}(Z_\mathrm{L}\teleportationchannel(Z_\mathrm{L}))&=\chi_{00}-\chi_{11}-\chi_{22}+\chi_{33}.
\end{align}
Therefore,
\begin{align}
    \frac12\left(\frac{1}{2}\mathrm{tr}(X_\mathrm{L}\teleportationchannel(X_\mathrm{L})) + \frac{1}{2}\mathrm{tr}(Z_\mathrm{L}\teleportationchannel(Z_\mathrm{L}))\right) &= \chi_{00}-\chi_{22} \\&\leq \chi_{00}\\&=F(\teleportationchannel)
\end{align}

where we used $0 \leq \chi_{22}$ implied by the positive semidefiniteness of the process matrix. Note that $\chi_{22}$ tends to zero as $\teleportationchannel$ tends to the identity, so the bound above is asymptotically tight when $\teleportationchannel\to\mathcal{I}$.

Because \cref{eq:average_fidelity} is an affine transformation of the entanglement fidelity, it implies a lower bound on the average fidelity of the channel.

\subsection{Bias of the Decoded $\langle\lopy\rangle$ in Teleported State Tomography}\label{app:state_teleportation_bias}

In the lattice surgery data (see \cref{fig:state_teleportation}f-i of the main text), we find that initializing the logical qubit 1 (L1) in the X-basis (to either \pluslogical{} or \minuslogical) and measuring in the Y-basis results in a non-zero (decoded) expectation value. This is unexpected because the teleported logical state does not commute with the logical measurement basis. Consequently the logical outcome in an ideal experiment should be fully random, such that $\langle\lopy{}\rangle \approx 0$. However, the neural-network decoder detects a bias in the logical measurement, leading to $\langle\lopy{}\rangle \approx 0.2$.

The exact source of the bias remains to be investigated, although we hypothesize that it is was caused by a systematic Z-rotation of all the data qubits in the initialization cycle (before the merge operation) that was uncompensated in these two circuits specifically.  In particular, in our matched basis experiments (i.e. experiments with a deterministic outcome such as measuring \pluslogical{} in the X basis), we calibrate virtual phases to compensate for the Z-rotations. However, these phases were not calibrated in the non-deterministic mismatched basis circuits (e.g.~measuring \pluslogical{} in the Y basis). 

We expect to be able to correct this phase in future work by performing an appropriate virtual-Z rotation on each data qubit.

\bibliography{resources/references}

%% file: preamble.tex
\usepackage{amsmath}    
\usepackage{graphicx}   
\usepackage{color}      
\usepackage[usenames,dvipsnames,svgnames,table]{xcolor}
\usepackage[hidelinks]{hyperref}   
\usepackage{braket}
\usepackage{bbm}
\usepackage[T1]{fontenc}

\usepackage[capitalise]{cleveref}
\Crefname{figure}{Fig.}{Figs.}
\Crefname{table}{Tab.}{Tabs.}
\usepackage{siunitx}
\usepackage{amssymb}

\usepackage{rotating}
\usepackage{tabularx}
\usepackage{cellspace}
\usepackage{booktabs}
\usepackage{graphics}
\usepackage{newfloat}
\usepackage{float}
\newcolumntype{Y}{>{\centering\arraybackslash}X}
\newcolumntype{d}{c}
\newcolumntype{a}{c}

\usepackage{xr}

\renewcommand{\i}{\mathrm i}

\newcommand{\auxiliary}{auxiliary}

\newcommand{\zerological}{\ensuremath{\ket{0_\mathrm{L}}}}
\newcommand{\onelogical}{\ensuremath{\ket{1_\mathrm{L}}}}
\newcommand{\pluslogical}{\ensuremath{\ket{+_\mathrm{L}}}}
\newcommand{\minuslogical}{\ensuremath{\ket{-_\mathrm{L}}}}
\newcommand{\magicstateA}{\ensuremath{\ket{A_\mathrm{L}}}}
\newcommand{\magicstateT}{\ensuremath{\ket{T_\mathrm{L}}}}
\newcommand{\magicstateH}{\ensuremath{\ket{H_\mathrm{L}}}}
\newcommand{\arbitrarymagicstate}{\ensuremath{\ket{\psi_\mathrm{L}}}}

\newcommand{\nrclifford}{\ensuremath{m}}

\newcommand{\lopz}{\ensuremath{\hat{Z}_{\mathrm{L}}}}
\newcommand{\lopy}{\ensuremath{\hat{Y}_{\mathrm{L}}}}
\newcommand{\lopx}{\ensuremath{\hat{X}_{\mathrm{L}}}}
\newcommand{\lopxx}{\ensuremath{\hat{X}_{\mathrm{L1}}\hat{X}_{\mathrm{L2}}}}
\newcommand{\px}{\ensuremath{\hat{X}}}
\newcommand{\pz}{\ensuremath{\hat{Z}}}

\newcommand{\syndrome}{\ensuremath{\sigma}}
\newcommand{\pdet}{\ensuremath{p_\mathrm{d}}}

\newcommand{\epc}[1][d]{\ensuremath{\varepsilon_{#1}}}
\newcommand{\meanepc}[1][d]{\ensuremath{\overline{\varepsilon}_{#1}}}

\newcommand{\nmemcycles}{\ensuremath{n}}
\newcommand{\errorsuppressionfactor}{\ensuremath{\Lambda_{3/5}}}
\newcommand{\errorsuppressionfactorsim}{\ensuremath{\Lambda_{3/5}^{\mathrm{sim}}}}

\newcommand{\logicalerrorprob}{\ensuremath{p_\mathrm{L}}}

\newcommand{\errorperclifford}{\ensuremath{\varepsilon_{\mathrm{C}}}}

\newcommand{\polarangle}{\ensuremath{\theta}}
\newcommand{\azimuthalangle}{\ensuremath{\phi}}

\newcommand{\teleportationstate}{\ensuremath{\ket{\Psi_\mathrm{L}}}}
\newcommand{\teleportationmxxoutcome}{\ensuremath{m_1}}
\newcommand{\teleportationmzoutcome}{\ensuremath{m_2}}
\newcommand{\mxx}{\ensuremath{\mathrm{M}_{XX}}}

\newcommand{\teleportationchannel}{\ensuremath{\mathcal{T}}}
\newcommand{\teleportationaveragefidelity}{\ensuremath{\overline{F}}}

\newcommand{\lambdaneuralnetworkvalue}{1.56(4)} 
\newcommand{\lambdabest}{\lambdaneuralnetworkvalue}
\newcommand{\lambdaneuralnetworksimvalue}{1.680(4)}
\newcommand{\epcfivebestvalue}{0.0110(2)} 
\newcommand{\meanepcthreebestvalue}{0.0171(3)} 
\newcommand{\pdetweightfourdistancefiveX}{0.112}
\newcommand{\pdetweightsixdistancefiveX}{0.149}

\newcommand{\errorpercliffordneuralnetworkvalue}{0.0027(3)} 

\newcommand{\polarangleinfidchromobiusvalue}{0.039(3)}
\newcommand{\polarangleinfidpostselectionvalue}{0.009(4)}
\newcommand{\polaranglepostselectionfraction}{74.8\,\%}

\newcommand{\magicstateAinfidpostselectionvalue}{$0.0008_{-3}^{+15}$} 
\newcommand{\magicstateHinfidpostselectionvalue}{$0.0041_{-38}^{+37}$} 
\newcommand{\magicstateTinfidpostselectionvalue}{$0.0086_{-65}^{+67}$} 
\newcommand{\magicstateAfullpostselectionfraction}{75.2\,\%} 
\newcommand{\magicstateHfullpostselectionfraction}{74.6\,\%} 
\newcommand{\magicstateTfullpostselectionfraction}{75.7\,\%} 

\newcommand{\teleportationfidelityneuralnetworkvaluemin}{86.5(1)\,\%}
\newcommand{\teleportationfidelityneuralnetworkvaluemax}{90.7(1)\,\%}
\newcommand{\teleportationentanglementfidelityneuralnetworkvalue}{84.7(1)\,\%} 

\newcommand{\Nathan}{N.L.}
\newcommand{\AlexandreB}{A.~Bourassa}
\newcommand{\Craig}{C.G.}
\newcommand{\KevinS}{K.J.S.}
\newcommand{\Matt}{M.M.}
\newcommand{\Cody}{C.J.}
\newcommand{\AndreasB}{A.~Bengtsson}
\newcommand{\Jimmy}{Z.C.}
\newcommand{\Alexis}{A.M.}
\newcommand{\Johannes}{J.B.}
\newcommand{\Francisco}{F.J.H.H.}
\newcommand{\Lei}{L.M.Z.}
\newcommand{\Noah}{N.S.}
\newcommand{\Adam}{A.Z.}
\newcommand{\Vlad}{V.S.}
\newcommand{\Oscar}{O.H.}
\newcommand{\Dvir}{D.K.}
\newcommand{\Jahan}{J.C.}
\newcommand{\Julian}{J.K.}

\newcommand{\puttitle}{Scaling and logic in the color code on a superconducting quantum processor}

\newcommand{\optional}[1]{{\color{black}{#1}}}
\newcommand{\ks}[1]{}
\newcommand{\ab}[1]{}
\newcommand{\nl}[1]{}

\newif\ifshowsectionnames

\showsectionnamestrue

%% file: authors_short.tex
\onecolumngrid

\vspace{-1.0em}
\begin{flushleft}

\bigskip
{\small\renewcommand{\author}[2]{#1$^\textrm{\scriptsize #2}$}
\renewcommand{\affiliation}[2]{$^\textrm{\scriptsize #1}$ #2 \\}

\newcommand{\xGoogle}{\affiliation{1}{Google Research, Mountain View, CA, USA}}

\newcommand{\xETHZurich}{\affiliation{2}{Department of Physics, ETH Zurich, Switzerland}}

\newcommand{\xGDM}{\affiliation{3}{Google DeepMind, London, England, UK}}

\newcommand{\xUMass}{\affiliation{4}{Department of Electrical and Computer Engineering, University of Massachusetts, Amherst, MA, USA}}

\newcommand{\xStorrs}{\affiliation{5}{Department of Physics, University of Connecticut, Storrs, CT, USA}}

\newcommand{\xUCSBCS}{\affiliation{6}{Department of Computer Science, University of California, Santa Barbara, CA, USA}}

\newcommand{\xMITLab}{\affiliation{7}{Research Laboratory of Electronics, Massachusetts Institute of Technology, Cambridge, MA, USA}}

\newcommand{\xMITEE}{\affiliation{8}{Department of Electrical Engineering and Computer Science, Massachusetts Institute of Technology, Cambridge, MA, USA}}

\newcommand{\xMITPhysics}{\affiliation{9}{Department of Physics, Massachusetts Institute of Technology, Cambridge, MA, USA}}

\newcommand{\xEqualContrib}{\affiliation{\textdaggerdbl}{These authors contributed equally to this work.}}

\newcommand{\Google}{1}
\newcommand{\ETHZurich}{2}
\newcommand{\GDM}{3}
\newcommand{\UMass}{4}
\newcommand{\Storrs}{5}
\newcommand{\UCSBCS}{6}
\newcommand{\MITLab}{7}
\newcommand{\MITEE}{8}
\newcommand{\MITPhysics}{9}
\newcommand{\EqualContrib}{\textdaggerdbl}

\author{N.~Lacroix}{\Google,\! \ETHZurich,\! \EqualContrib},
\author{A.~Bourassa}{\Google,\! \EqualContrib},
\author{F.~J. H.~Heras}{\GDM},
\author{L.~M. Zhang}{\GDM},
\author{J.~Bausch}{\GDM},
\author{A.~W.~Senior}{\GDM},
\author{T.~Edlich}{\GDM},
\author{N.~Shutty}{\Google},
\author{V.~Sivak}{\Google},
\author{A.~Bengtsson}{\Google},
% ...

\bigskip

\xGoogle
\xETHZurich
\xGDM
\xUMass
\xStorrs
\xUCSBCS
\xMITLab
\xMITEE
\xMITPhysics

\smallskip

\xEqualContrib
}
\end{flushleft}

\clearpage
\twocolumngrid